\documentclass[runningheads,a4paper]{llncs}
\usepackage{cite}
\usepackage{amssymb}
\usepackage{graphicx}
\usepackage{amsmath,amssymb,amsfonts}
\usepackage{url}
\urldef{\mailsa}\path|{alfred.hofmann, ursula.barth, ingrid.haas, frank.holzwarth,|
\urldef{\mailsb}\path|anna.kramer, leonie.kunz, christine.reiss, nicole.sator,|
\urldef{\mailsc}\path|erika.siebert-cole, peter.strasser, lncs}@springer.com|    

\newcommand\be{\begin{equation}}
\newcommand\ee{\end{equation}}
\newcommand\la{\langle}
\newcommand{\ra}{\rangle}

\begin{document}

\mainmatter  

\title{Superoscillations and Physical Applications}


%
%
\author{Andrew N. Jordan\inst{1,2} \and John C. Howell \inst{2}\and Nicholas Vamivakas\inst{3} \and Ebrahim Karimi\inst{4,2}}

\institute{The Kennedy Chair in Physics, Chapman University, Orange, CA 92866, USA
\and
Institute for Quantum Studies, Chapman University, Orange, CA 92866, USA
\and
The Institute of Optics, University of Rochester, 480 Intercampus Dr, Rochester, NY 14627, USA
\and
Department of Physics, University of Ottawa, 25 Templeton, Ottawa, K1N 6N5, ON, Canada}

\authorrunning{Jordan {\it et al.}: Superoscillations and Physical Applications}


%
%

\maketitle

\begin{abstract}
This book chapter gives a selective review of physical implementations and applications of superoscillations and associated phenomena. We introduce the field by reviewing simple examples of superoscillations and showing how their existence naturally follows from the real part of the quantum mechanical weak value, which the parallel phenomena of supergrowth naturally follows from the imaginary part.  Focusing on electromagnetic applications, we review the topics of superoscillation and supergrowth in speckle, creating superoscillating hot spots with patterned filters, superspectroscopic discrimination of two molecules, noise mitigation and the engineering of super behavior in point spread functions for the purpose of optical superresolution.  We also cover a variety of different methods for creating superoscillatory and supergrowing functions, reviewing both mathematical and physical ways to create this class of functions, and beyond.  Promising directions for future research, including superoscillations in other wave phenomena, super radar, and generalized super-phenomena in quantum physics, are outlined.
\end{abstract}

\section{Introduction}
Superoscillation is a phenomenon that was independently discovered by Yakir Aharonov et al. \cite{aharonov1847can}, and Michael Berry \cite{berry1994evanescent}.
The basic effect is to consider functions that are {\it band-limited}, that is functions whose Fourier transform is strictly zero above a highest frequency, and yet nevertheless oscillate {\it locally} with a frequency that is faster than the highest frequency. 

Aharonov {\it et al} considered a simple example \cite{aharonov2011some} of a function $f$ given by
\be
f(x) = \left[ \cos(x/N) + i a \sin(x/N) \right]^N.  \label{f}
\ee
by making a binomial expansion, the function is converted into a Fourier series,
\be
f(x) = \sum_{n=0}^N c_n e^{i k_n x},
\ee
with wavenumbers $k_n  = 1 - 2n/N$ and Fourier coefficients 
\be
c_n = \begin{pmatrix} N \\ n \end{pmatrix} ((1 + i a)/2)^{N-n}  ((1 - i a)/2)^{n}.
\ee
We note that the highest frequency is $|k_{max}| = 1$.

Nevertheless, by making a Taylor expansion of $f$ around $x=0$, we note that
\be
f(x) \approx 1 + i a x + {\cal O}(x^2) \approx e^{i a x}.
\ee
Thus, in the vicinity of $x=0$, the function locally oscillates with a frequency of $a$.  Since $a$ is a parameter, this oscillation rate can be as large as desired - however, we note that the price to be paid for this local high oscillation frequency is the exponentially growing values of the function away from $x=0$ as a function of increasing local frequency.

While various formal definitions of superoscillations can be mathematically formulated, a simple and appealing one is whenever the local wavenumber exceeds the bandlimit of the function.  In the case of the above example, the band limits are $k=\pm 1$.  The local wavenumber is defined as
\be
k(x) = {\rm Im} [\partial_x \ln f(x)]. 
\label{localk}
\ee
Applied to the above example, the local wavenumber becomes $a$ at $x=0$, which exceeds 1, thus qualifying for a superoscillatory region.

Historically, Aharonov et al. found this type of function via the theory of the quantum mechanical weak value \cite{PhysRevLett.60.1351,dressel2014colloquium,jordan2024quantum}.  We may view the above criterion for the superoscillation region as identical to the anomalous weak value of the momentum operator ${\bf p} = i \partial_x$, where we have set the reduced Planck constant to 1.  Given a pre-selected wavefunction $f(x) 
= \la x | f\ra$, where we use the Dirac inner product notation, and a post-selected position $\la x |$, we have the weak value of the momentum operator to be
\be
p_w = \frac{\la x | {\hat p} | f \ra}{\la x | f \ra} = i \partial_x f(x) / f(x).
\ee
Thus, we see that taking the {\it real part} of the weak value gives the above definition of the local wave number.  The weak value is defined to be anomalous when the weak value exceeds the largest eigenvalue of the operator $\bf p$.  This anomalous weak value definition coincides with the definition of the superoscillation in this special case. 

\begin{figure}
\centering
\includegraphics[height=5.5cm]{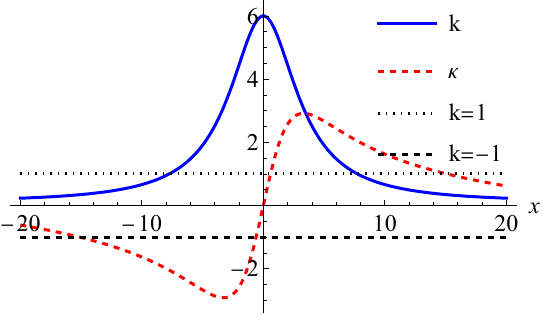}
\caption{The local wavenumber $k$, Eq.~(\ref{localk}) and the local growth rate $\kappa$, Eq.~(\ref{localkappa}) are plotted versus $x$.  The band limits $k = \pm 1$ are shown as dashed lines.  Here we choose $N=20, a=6$.  Regions of superoscillation and supergrowth are when the functions stray outside the dashed boundaries, and are generally different.  In this example the point of maximum superoscillation at $x=0$ has zero supergrowth.  }
\label{fig:ks}
\end{figure}

It is interesting to see that the imaginary part of the weak value has an important role to play in quantum mechanics.  It is therefore a natural question if the real part of the logarithmic derivative of $f(x)$ has any special role.  Indeed, if superoscillation is the variation of the local phase faster than the highest frequency in the Fourier series, then we define {\it supergrowth} as the variation of the local amplitude faster than the highest frequency \cite{jordan2020superresolution}.  We formalize this definition as the imaginary part of the weak value, or as the real part of the logarithmic derivative,
\be
\kappa(x) = {\rm Re} [\partial_x \ln f(x)].
\label{localkappa}
\ee
Whenever $\kappa$ exceeds the bandlimit, the function is said to be supergrowing (or decaying) at that point.  Both the local growth rate and the local wavenumber of the function $f(x)$, Eq.~(\ref{f}) are plotted in Fig.~\ref{fig:ks}.  

The focus of the current article is the physical implementations and applications of superoscillation and supergrowth.  While these concepts can be applied to any wave phenomena, including water waves, sound waves, and quantum matter waves, we focus in this selective review on the area where there has been the most intense research:  electromagnetic waves.  From the above description of the mathematical foundation of superoscillations, its main feature is that the function appears to oscillate faster than the highest Fourier frequency.  This gives rise to a number of interesting physical effects that we will discuss in detail below.
Perhaps the first physical effect described that could be classified as a superoscillatory effect is the phenomenon of highly directive antennas \cite{schelkunoff1943mathematical}.  In that proposal, the radiation emitted from an antenna is directed into a solid angle that is much smaller than is expected from a usual dipole antenna \cite{cox1986practical,haviland1995supergain}.  We refer the interested reader to the summary of linear antenna arrays, super-gain antennas, and endfire in Berry's book chapter \cite{berry2014superoscillations}.

We begin our review in Sec.~\ref{influential} where we make a selective treatment of a collection of experiments designed to highlight increasing complexity and control of superoscillatory and supergrowing phenomenon.  

In Subsection \ref{speckle}, we begin our review by stressing that superoscillatory effects are neither special nor rare.  They occur quite generically in random waves \cite{berry2008natural}.  Indeed, they occur ``naturally'' in speckle patterns \cite{dennis2008superoscillation}, occupying a fraction as large as a third of the spatial region.  Speckle patterns are random optical waves formed by propagation through random media.  Although these superoscillations are weak in strength (deviation above the band limit), they are stable upon propagation, which is quite different from designed superoscillation behaviors \cite{berry2006evolution}, which showed that superoscillatory structures can be fragile to propagation.

In subsection \ref{hotspots}, we review the experiment of Huang {\it et al.} \cite{huang2007optical}.  They showed that an optical mask consisting of quasi-periodic holes can create local ``hot-spots'' of superoscillatory regions.  More generally, the mask forms ``carpets'' of patterns that have structures that can be much smaller than the wavelength of the light that is used to illumuinate the mask.  The important step forward from a random medium is the structure in the placement of the holes in the mask.


In subsection \ref{super-spectroscopy} we discuss how superoscillations can enable super-resolution spectroscopy in the THz range. McCaul et al. \cite{mccaul2023superoscillations} employed a novel method to generate THz waveforms, differing from the typical approach of fitting a waveform to a specific function using periodic frequencies. Instead, they were constrained to four quasi-cw THz frequencies, which arise from the optical rectification of a 1030 nm femtosecond laser in periodically poled lithium niobate.

In subsection \ref{psfengineer} the experiment of Kozawa {\it et al.} \cite{kozawa:18} demonstrates the ability to {\it engineer} the point spread function of an optical imaging system to obtain superoscillatory behavior.  The benefit of this system is that higher resolution can be obtain in microscopy applications.  Indeed, the authors show that biological systems can be imaged more sharply with this system than could be obtained with conventional imaging using the shortest wavelength of the light they used.  This fact circumvents the Rayleigh criterion that limits the resolution of a conventional imaging system like microscopes or telescopes.

In subsection \ref{supergrowth}, we describe the dual manifestation of superoscillation - supergrowth, and how it can be observed and used in optical systems.  The experiment of Sethuraj {\it et al.} \cite{kr2024experimental}
shows that supergrowth gives the same resolution improvements as superoscillation but has the advantage of having much more irradiance in the supergrowing region.  This validates the prediction that supergrowth can have much better signal to noise performance in optical superresolution \cite{jordan2020superresolution}.

Subsection \ref{recovering}, shows how to overcome a common criticism of superoscillation methods - namely that these structures have a very small intensity, and are thus easily swamped by noise in any realistic application.  The experiment of White {\it et al.} in Ref.~\cite{white2024reconstructing} shows that superoscillations can be recovered even in high noise environments.  When superoscillatory behavior is encoded in a set of discrete frequencies, filtering techniques can restore the superoscillation even when the noise is orders of magnitude larger in amplitude than the superoscillatory structures.

Having covered some of the key experiments in the field, we then turn to different methods of how to construct superoscillatory patterns that can then be applied in experiments in Sec.~\ref{methods}.  A variety of different constructive approaches have been discovered over the years, and we cover the main ones.

We conclude our review in Sec.~\ref{conclusions} with a survey of topics that are adjacent to this review, such as superphenomena in quantum physics, as well as different types of waves that can support superoscillations, such as acoustic waves. We also cover new developments such as super ranging resolution in radar (superradar), and give an outlook on the future of the field.

\section{Influential experiments and effects}
\label{influential}

\subsection{Random and pseudo-random waves} 
\label{speckle}

It is an understatement to say the advent and availability of visible lasers in the 1960's led to many surprises \cite{siegman:86}. One of these occurred when laser light was reflected off a white matte screen (a diffuse reflecting surface)\cite{oliver1963sparkling}. The observed reflected laser light exhibited, stable, randomly occurring bright and dark spots.  It was soon discovered that the origin of the observed pattern was the superposition of randomly phased reflected wavetrains and the observed phenomenon became known as speckle.  Fast forward to today and speckle has become a fascinating area of modern optics \cite{goodman2007speckle} with impact in technology such as sensing and imaging \cite{redding2012speckle, redding2013all, heeman2019clinical, luo2020parametrization,hastings2025localization} while at the same time continuing to provide opportunities for new fundamental understanding in optics.  One example of the latter was to understand if speckle supported superoscillatoins and the nature of such wave behavior \cite{dennis2008superoscillation}.  More recently, the supergrowing behavior in speckle has been theoretically described and experimentally anlayzed \cite{viteri2024supergrowth}.  

The theoretical work of Dennis and coauthors set out with the explicit goal of ``to study some simple superoscillatory aspects of two-dimensional random waves" \cite{dennis2008superoscillation}.    The intuition driving the work was the recognition that optical vortices, locations of undefined phase in a beam or field, exhibit divergent phase at the intensity null and should possess superoscillations.  The analysis was focused on if such behavior may be broadly engaged across a given plane of speckle and extends prior work on vortices in random fields and speckle \cite{berry1978disruption, baranova1981dislocations, berry2000phase}.  The main theoretical result begins with assuming waves of the form $\psi = \rho e^{i\chi}$ that have intensity $I=\rho^2$ and phase $\chi$ and deriving the following joint probability distribution for the intensity $I$ and local phase variation $|\nabla \chi|$ [see Eq. (\ref{localk})]
 \begin{equation}
   P(I,|\nabla \chi|) = \frac{|\nabla \chi|}{I^2_ok_2}\text{exp}\left(-\frac{I}{I_o}(1+|\nabla \chi|^2/2k_2) \right).
       \label{Dennis1}
\end{equation}

\begin{figure}[hbt!] 
\centering
\includegraphics[width=0.5\textwidth]{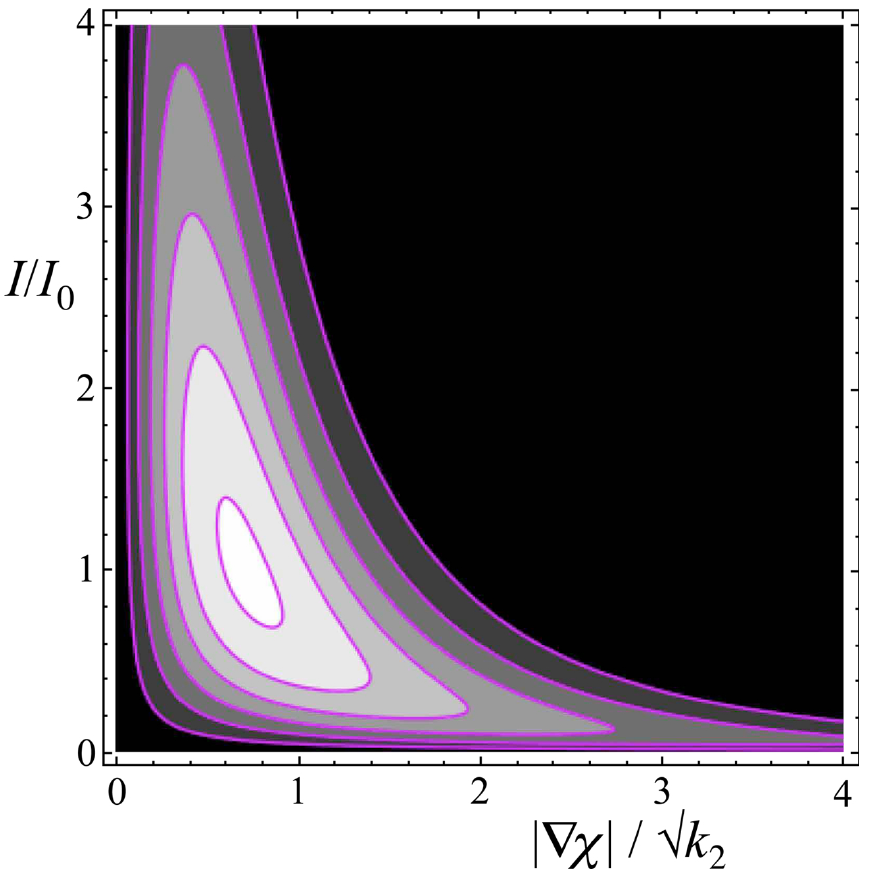}
\caption{\label{DenFig1} Visualization of the joint probability distribution presented in Eq. (\ref{Dennis1}).  Fast phase variation occurs in regions of small irradiance. Taken from  \cite{dennis2008superoscillation}. }
\end{figure}

\noindent In Eq. (\ref{Dennis1}), $I_o$ is the average intensity across the given plane and $k_2$ is the normalized second moment of the power spectrum.  In this work the power spectrum has a well defined $k_{max}$ and is circularly symmetric in $k$-space.  Figure \ref{DenFig1} illustrates the joint probability distribution and captures the expectation that locations of superoscillation correspond to places where the intensity is low.

\noindent By integrating Eq. (\ref{Dennis1}) over $I$ the probability density of $|\nabla\chi|$ is found:
 \begin{equation}
   P(|\nabla \chi|) = \frac{4k_2|\nabla \chi|}{(2k_2+|\nabla \chi|^2 )^2}
       \label{Dennis2}
\end{equation}
\noindent and could be used to determine the fractional amount, $f=\int_{k_{max}}^{\infty}d|\nabla\chi|P(|\nabla\chi|)$, of supergrowing area in the speckle.  The result of integrating Eq. (\ref{Dennis2}) is $f=1/3$ assuming an annular $k$-space spectrum and $f=1/5$ for a disk spectrum.

The work of Dennis inspired recent work by Viteri-Pflucker and collaborators \cite{viteri2024supergrowth} to explore the supergowth phenomena in speckle fields. The work was part theoretical, comparing the results of supergrowth and superoscillation for a disk spectrum. And part experimental, analyzing speckle fields in the lab to characterize their supergrowing behavior.   Since supergrowth is an amplitude variation phenomena, it is not necessary to measure the full-field to observe it in speckle, the irradiance carries its signature. Theoretical analysis in \cite{viteri2024supergrowth} confirmed, for the disk spectrum, supergrowth occupied a fractional area of $1/5$ across the speckle plane.   

\begin{figure}[hbt!] 
\centering
\includegraphics[width=0.75\textwidth]{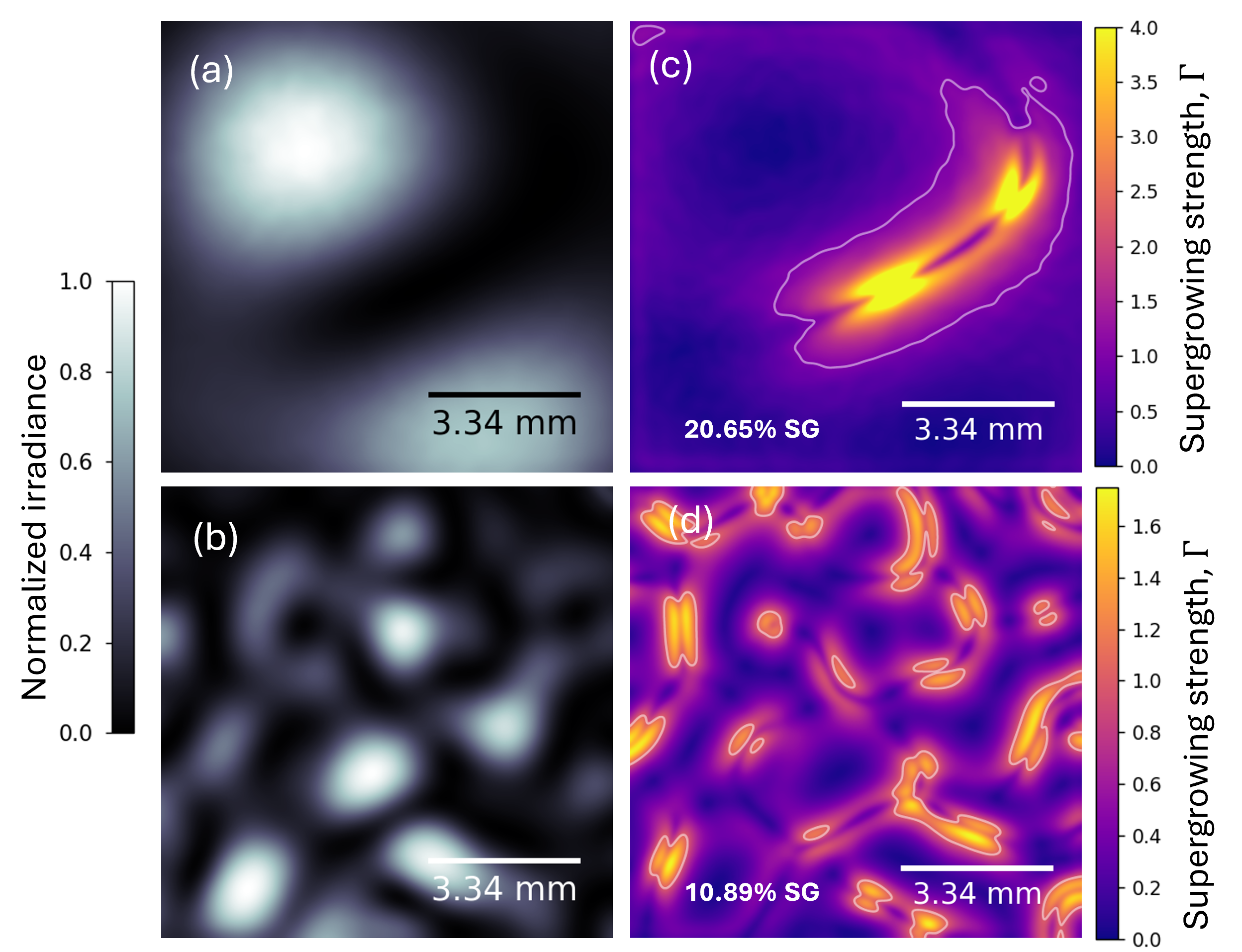}
\caption{\label{ValFig1} (a), (b) Two different speckles realizations.  The speckle is created by passing a focused laser through ground glass. The laser beam size on the ground glass controls the observed speckle structure.  (c),(d) The observed supergowth in the speckle irradiance of panel (a), (b) where the growth parameter $\Gamma>1$ indicates local amplitude variation exceeding that predicted by the bandlimit. Taken from Ref.~\cite{viteri2024supergrowth}.}
\end{figure}

Figure  \ref{ValFig1}(a) and (b) presents two speckle realizations generated by passing a focused laser though ground glass.  The ground glass illumination area is different for Fig. \ref{ValFig1}(a) and (b) so the corresponding speckle irradiance patterns are not the same.  Figure \ref{ValFig1}(c) and (d) present the supergrowth structure intrinsic to the speckle irradiance.   The supergrowing strength parameter $\Gamma$ captures by how much locally the amplitude variation is larger than the bandlimit.  $\Gamma>1$ is indicative of supergrowth. Figures \ref{ValFig1}(c) and (d) provide experimental evidence that supergrowth is indeed an intrinsic feature of speckle.

The previous discussion assumed random scalar waves superposed in 2-dimensions.  It is natural to extend notions of superoscillation and supergrowth to 3-dimensional scalar waves. In this case the random wave superposition is in a volume and not across a plane. Superocillating characteristics of random scalar wave superpositions, for any dimension $D$, were described in \cite{berry2008natural}. The quantity $f$, described after Eq. (\ref{Dennis2}), was found to to be $\approx 0.35$ - larger than the $f=1/3$ found for two-dimensions.  The situation becomes even richer if one considers vector waves in multiple dimensions. Such an analysis was carried out by Berry and coauthors in \cite{berry2019geometry}.  Experimental confirmations of these more general cases, as well as their suprgrowing behaviors, are an open area of future research. 

\subsection{Hot-Spots}
\label{hotspots}
In \ref{speckle}, we have described how the interplay between randomness, in the ground glass, and coherence, of the illuminating wave, lead to speckle distributions that contain vortices that are accompanied by areas of superoscillation and supergrowth.   The work by Huang and collaborators \cite{huang2007optical} studied how coherence and pseudorandomness could lead to down stream field distributions that exhibit superoscillations.  Particularly interesting in their work is their observation of local hot spots at different distances from a transmissive pseudorandom hole array. The holes were created via electron beam lithography in a 100-nm thick aluminum film on silica.   The hole array had approximately 10-fold symmetry with 14000 holes, each with a 200-nm diameter.  The sample diameter was 0.2-mm and the minimum hole separation was 1.2-$\mu$m.  The observed diffraction patterns are the pseudorandom counterpart of the talbot effect and were termed photonic carpets by the authors of \cite{huang2007optical}.  

\begin{figure}[hbt!]
\centering
\includegraphics[width=0.6\linewidth]{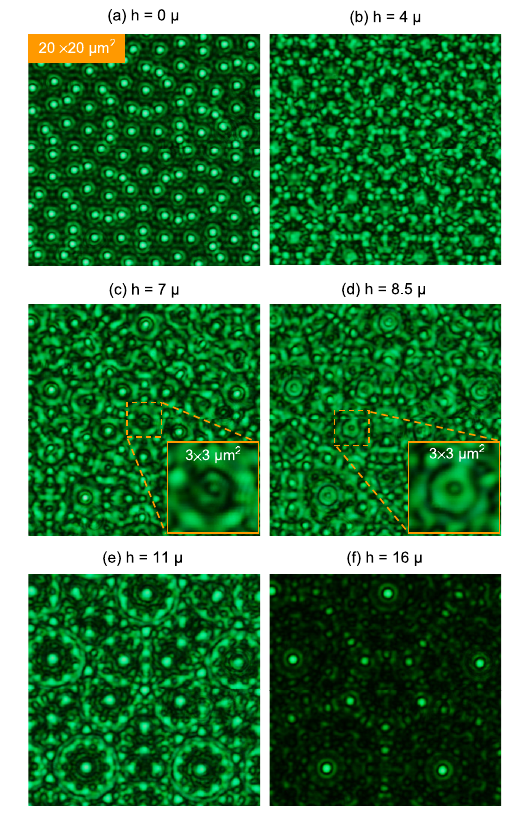}
\caption{\label{PhoCar_fig1} Observed photon carpet diffraction patterns at different distances from the hole array.  The middle two panel insets show local hotspots.  The field-of-view is $\sim$20-$\mu$m by 20-$\mu$m. Taken from Ref.~\cite{huang2007optical}.} 
\end{figure}

Figure \ref{PhoCar_fig1}, taken from \cite{huang2007optical}, presents the observed diffraction patterns at nine different planes above the hole array.  An exotic irradiance distribution is observed.  In the middle two panels, two hot spots are identified within the inset.  Each hot spot has a width of $\sim$400-nm.  The illumination wavelength in this case is 500-nm.  Without any focusing optics, it is possible to observe light concentration via the random hole array.  These experiments demonstrated how superoscillations could be used to create superresolved point spread functions with application in high resolution optical imaging.

\subsection{Super-Spectroscopy}
\label{super-spectroscopy}
In 2023, McCaul \textit{et al.} showed that superoscillations could be used for super resolution spectroscopy in the THz regime \cite{mccaul2023superoscillations}. Their results are remarkable on several levels. McCaul \textit{et al.} used a novel superoscillation creation technique to generate their THz waveforms.  Rather than prescribe a waveform to fit a certain function over a region using periodic frequencies, they operated under a different constraint.  In their case, they had a resource of four fixed-frequency quasi-cw THz frequencies that resulted from optical rectification of a 1030 nm femtosecond laser in periodically poled lithium niobate. The resultant wave of combining the four frequencies would be of the form
\begin{equation}
    f(t)=\sum_i A_i e^{i\omega_i (t-\tau_i)}.
\end{equation}
In creating their superoscillatory waveform, they assumed that the amplitudes of the four waves were all equal, \textit{i.e.},  $A_i=A$ for all $i$.  They then minimized the interference function over an interval $[-T_{SO},T_{SO}]$ by varying the phases $\tau_i$ of the waves using techniques such as gradient descent
\begin{equation}
I(\{\tau_i\})= \int_{-T_{SO}}^{T_{SO}}\left[\Sigma_i Ae^{i\omega_i (t-\tau_i)}\right]^2
\end{equation}

\begin{figure}
    \centering
    \includegraphics[width=1\linewidth]{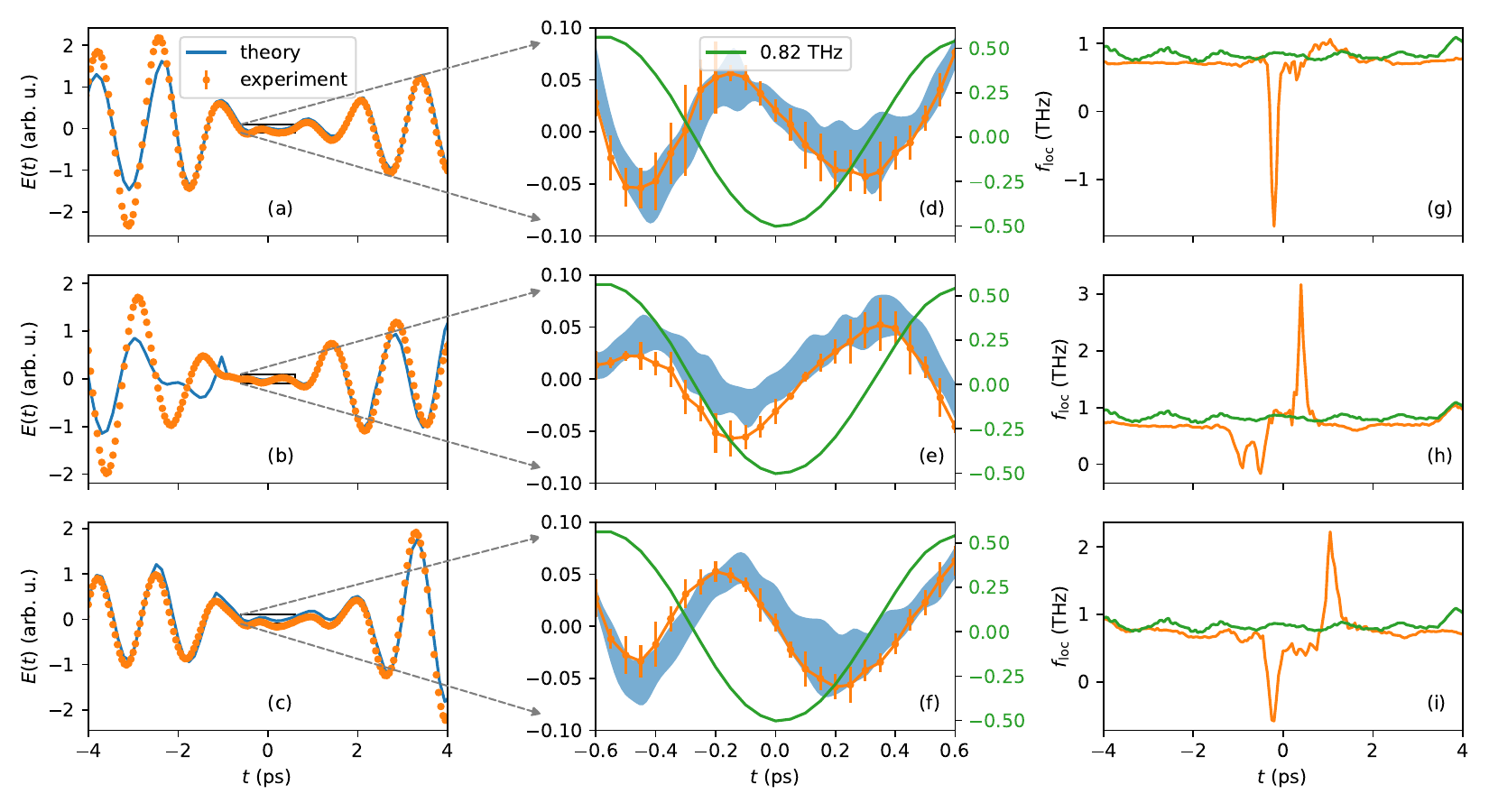}
    \caption{Three examples of superoscillation combinations. [(a)-(c)] show the predicted and measured result field over several oscillations, [(d)-(f)] show the superoscillation region comparing the highest frequency in the sum against the superoscillation frequency and [(g)-(h)] show the local frequencies of the largest wave and superoscillation. Reproduced from Ref.~\cite{mccaul2023superoscillations}.  Published by permission from APS.}
    \label{superSpecData}
\end{figure}

This technique does not guarantee superoscillations in the minimization region like other techniques.  However, they found that in tested cases, the resulting interference from the superposition of the four waves was superoscillatory. Their results are shown in Fig. \ref{superSpecData}

\begin{figure}
    \centering
    \includegraphics[width=.6\linewidth]{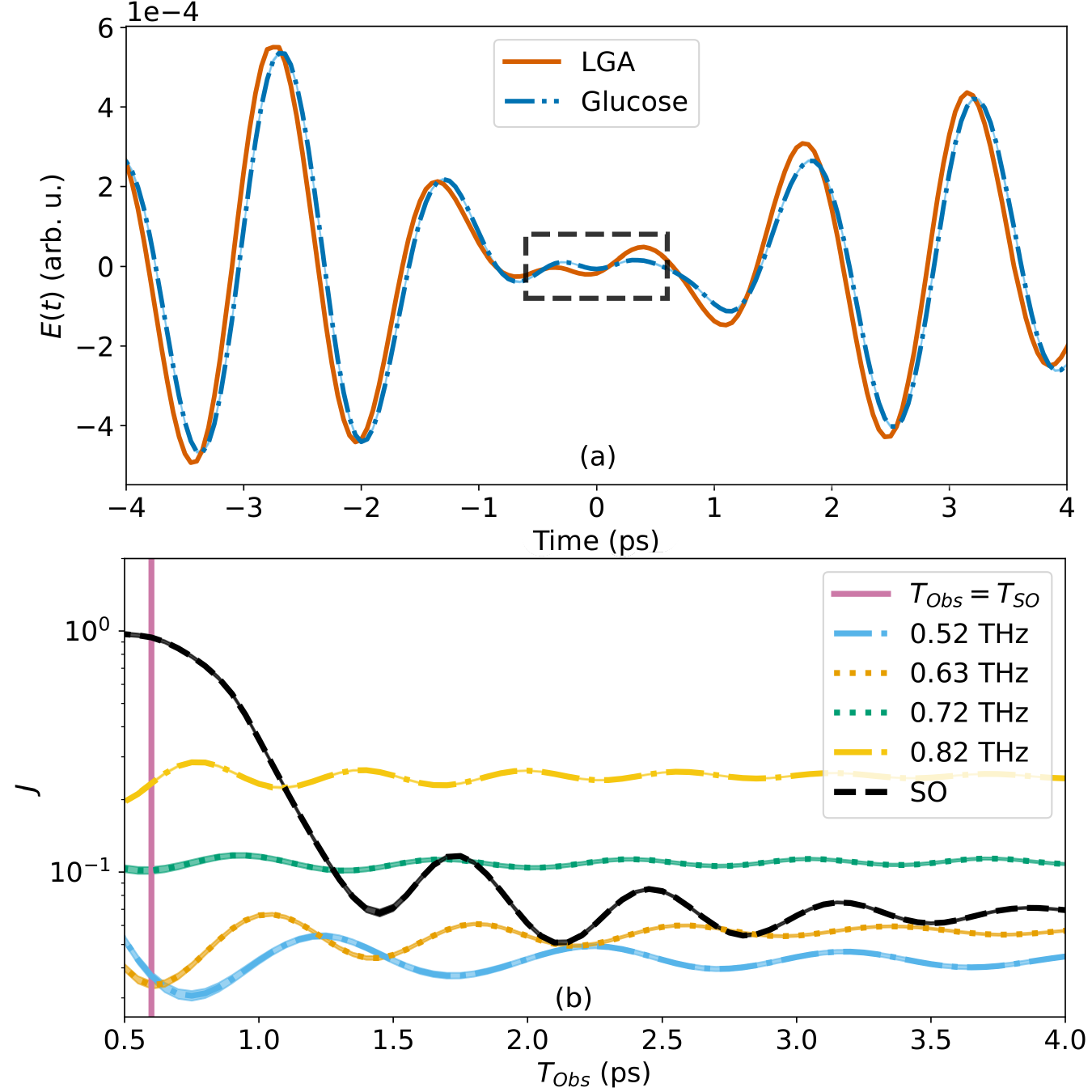}
    \caption{The top figure shows the differences in spectra between LGA and Glucose.  Clearly, in the superoscillatory region, there are significant differences in the signals, but very little distinction between the signals outside the region.  The Lower figure quantizes the discrimination as a function of position for the various waves compared to the superoscillating signal. Reproduced from Ref.~\cite{mccaul2023superoscillations}.  Published with permission from APS.}
    \label{superSpecDiscrimination}
\end{figure}

Once they crafted their superoscillatory waves, they then used the waveforms to probe the spectroscopic structure of $\alpha$-d-Glucose and L-Glutamic Acid, which have nearly identical absorption and refractive index profiles in the frequency band of the four frequencies used.  In the superoscillatory region, there is a clear separation in frequency structure between the two chemicals showing the resolving power of the system as shown in Fig. \ref{superSpecDiscrimination}.   

\subsection{Point spread function engineering}  \label{psfengineer}


Optical lenses are subject to a fundamental resolution limit dictated by both their geometrical parameters and the wavelength of the illuminated light beam. This limitation, commonly referred to as the Abbe or Rayleigh limit, defines the resolution of an optical system as $\lambda/(2\,\text{NA})$, where $\lambda$ is the illumination wavelength and NA is the numerical aperture of the system~\cite{goodman:05}. Essentially, the resolution is directly linked to the system’s ability to focus light tightly, which depends on the wavelength of light used and the lens’s numerical aperture, i.e., the highest \emph{wavevector}. To achieve higher resolution, shorter wavelengths of light or optical systems with a higher numerical aperture are logically preferred. However, increasing the numerical aperture (which is bounded) often requires immersing the lens in a medium like oil, which still has physical limitations. 
On the other hand, using shorter wavelengths of light increases the energy of the illumination, which can damage biological specimens, particularly living cells. These constraints have driven the development of alternative techniques to enhance resolution beyond the traditional limits of optical systems. Several advanced methods have emerged to address these challenges, including Stimulated Emission Depletion (STED) microscopy~\cite{STED:94}, super-resolution microscopy~\cite{tsang:16}, and super-oscillatory lenses~\cite{rogers:12}. Each approach has its own advantages and applications. In this section, we focus on the concept of super-oscillatory functions and their role in creating ``hot spots''--regions of highly localised light intensity. These hot spots enable resolution improvements by an order of magnitude, significantly surpassing conventional diffraction limits while maintaining compatibility with existing optical systems and wavelengths. While some debates have centered on linking STED microscopy to superoscillations near the phase singularity of a doughnut beam, the authors maintain that the technique is more closely associated with the nonlinear optical properties of the fluorescent dye and the size of the \emph{null} singularity. Although several review articles discuss the concept and applications of super-oscillating functions in improving the resolution of optical systems, see e.g.~\cite{gbur2019using, chen:19, zheludev:22}, here, as a matter of consistency, we only focus on a few ideas and original implementations.

Like many other fundamental original proposals, there have been debates on whether superoscillating functions remain theoretical concepts or will find applications in technologies. Sir Michael Berry and Sandu Popescu, in 2006, proposed a theoretical framework suggesting that diffraction from periodic structures can give rise to intricate, non-evanescent wave, speckle patterns, within which certain regions exhibit sub-wavelength intensities that surpass the conventional diffraction limit~\cite{berry2006evolution}. Their analysis demonstrated that these sub-wavelength structures are not merely localised phenomena but persist over an extended propagation range, challenging the previous notions of super-oscillating (wave) functions.

Building upon Berry \& Popescu’s theoretical proposal, experimental verification was later accomplished through a precisely designed quasicrystal structure exhibiting a tenfold symmetry~\cite{huang:07}. This was achieved by fabricating approximately 14,000 holes, each with a diameter of 200 nm, in an aluminium-based material. The resulting structure enabled the observation of the predicted sub-wavelength features--hot spots measuring $0.36 \lambda$ in size within a range of nearly $7 \lambda$--providing compelling empirical evidence that diffraction from complex periodic arrangements can yield finer optical features than the classical diffraction limit.

\begin{figure}
    \centering
    \includegraphics[width=1 \linewidth]{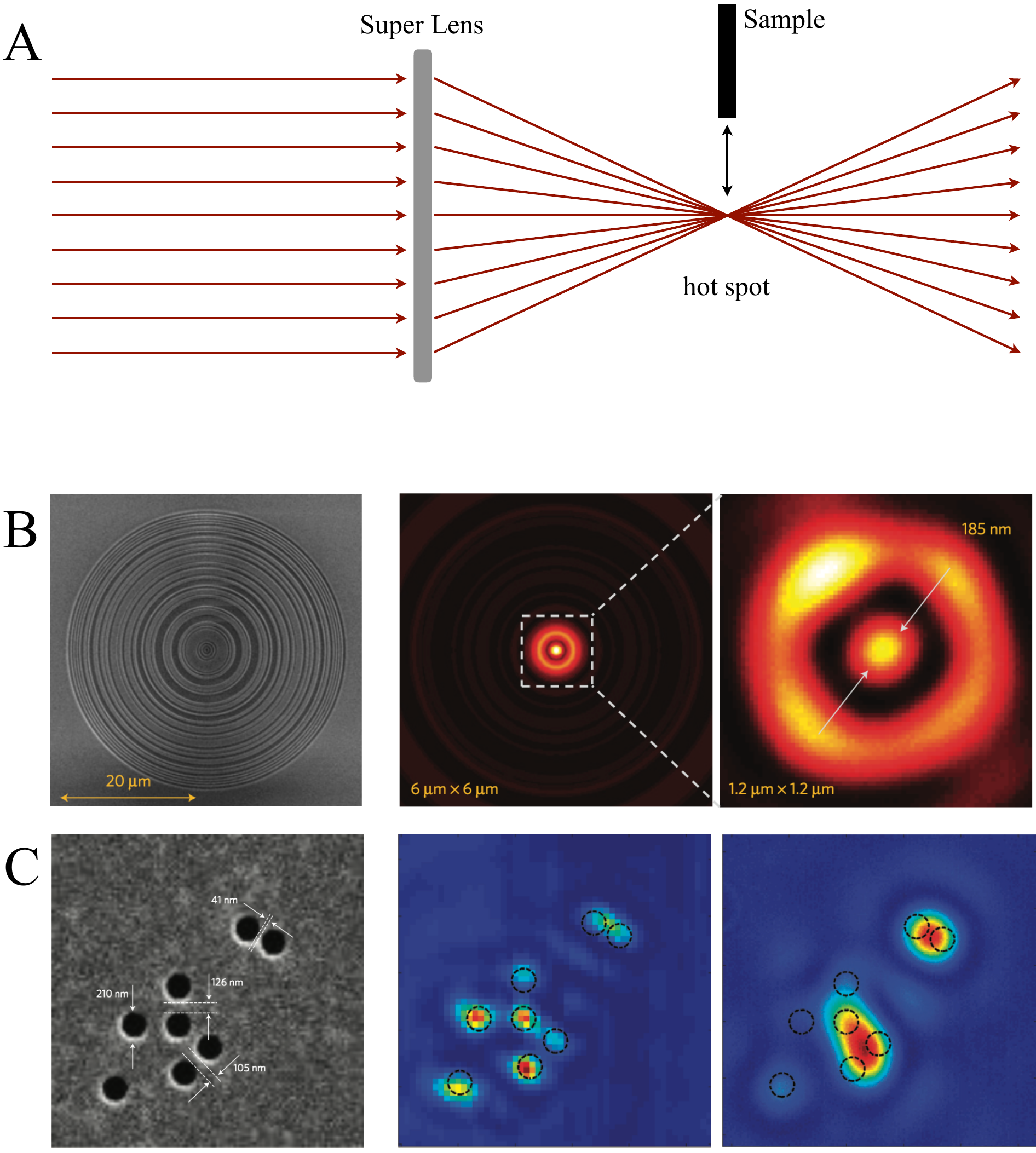}
    \caption{(A) Schematic configuration of an optical imaging system. The sample is placed at the hotspot, and the image is recorded subsequently. (B) Image of the designed super-oscillatory lens composed of concentric rings that allow light to either pass through or be blocked.  The outer ring diameter is about $40~\mu$m. The shown super-oscillatory lens is manufactured by focused ion-beam milling on a 100-nm-thick aluminium film on a glass substrate. The resulting intensity of the diffracted light at the focus, along with the zoomed-in image of the hotspot region, is presented. The SEM image of the random (cluster) nanoholes (with the dimension of 210 nm) on a metal film, along with their spacing distances (varying from 41 nm to 126 nm), are depicted in (C). As observed, the super-oscillatory lens can spatially resolve the random nanoholes, while the structure remains ‘hidden’ when illuminated with a Gaussian beam with an NA=1.4.  After Ref.~\cite{rogers:12}.}
    \label{figSL-exp}
\end{figure}

A few years later, several research teams developed precisely designed concentric nano-ring masks that enabled the selection of illuminated light within specific regions. The transmitted light after these masks produced a symmetric concentric diffraction pattern featuring a central hot spot beyond the diffraction limit~\cite{wang:10,rogers:12}. These nano-ring masks, referred to as super-oscillatory lenses, were binary amplitude masks, absorbing the illuminated light except for the transparent regions, thus reducing the transmitted power. The areas where the binary amplitude masks selected incoming beams were identified numerically and through proper diffraction simulation. These super-oscillatory lenses were used to focus the light beam beyond the diffraction limit at a hot spot in the far field and were employed to resolve nanostructures as well as a double slit successfully. A challenge these super-oscillatory lenses faced was the energy in the sidelobe region at the diffraction plane, which contains significantly more energy. The energy in the sidelobes posed challenges, as selecting the proper signal from a background of high noise was difficult. Though imaging with a super-oscillatory lens theoretically has no physical limits on resolution, it was experimentally shown that the hot spot created by the super-oscillatory lens could provide a better spatial resolution than a conventional microscope, namely, achieving a resolution surpassing $\lambda/6$. The super-oscillatory lens generated a hotspot of 185 nm (full width at half maximum) with a light wavelength of 640 nm at a distance of $10.4~\mu$ m in a conventional liquid immersion microscope with an NA of 1.4. Figure~\ref{figSL-exp} shows the configuration, image of the hot spot, and the obtained experimental results. Super-oscillatory lens can spatially resolve and provide images of random clusters of nanoholes, while conventional microscopy with a Gaussian beam is incapable of such a resolution.

\begin{figure}
    \centering
    \includegraphics[width=1 \linewidth]{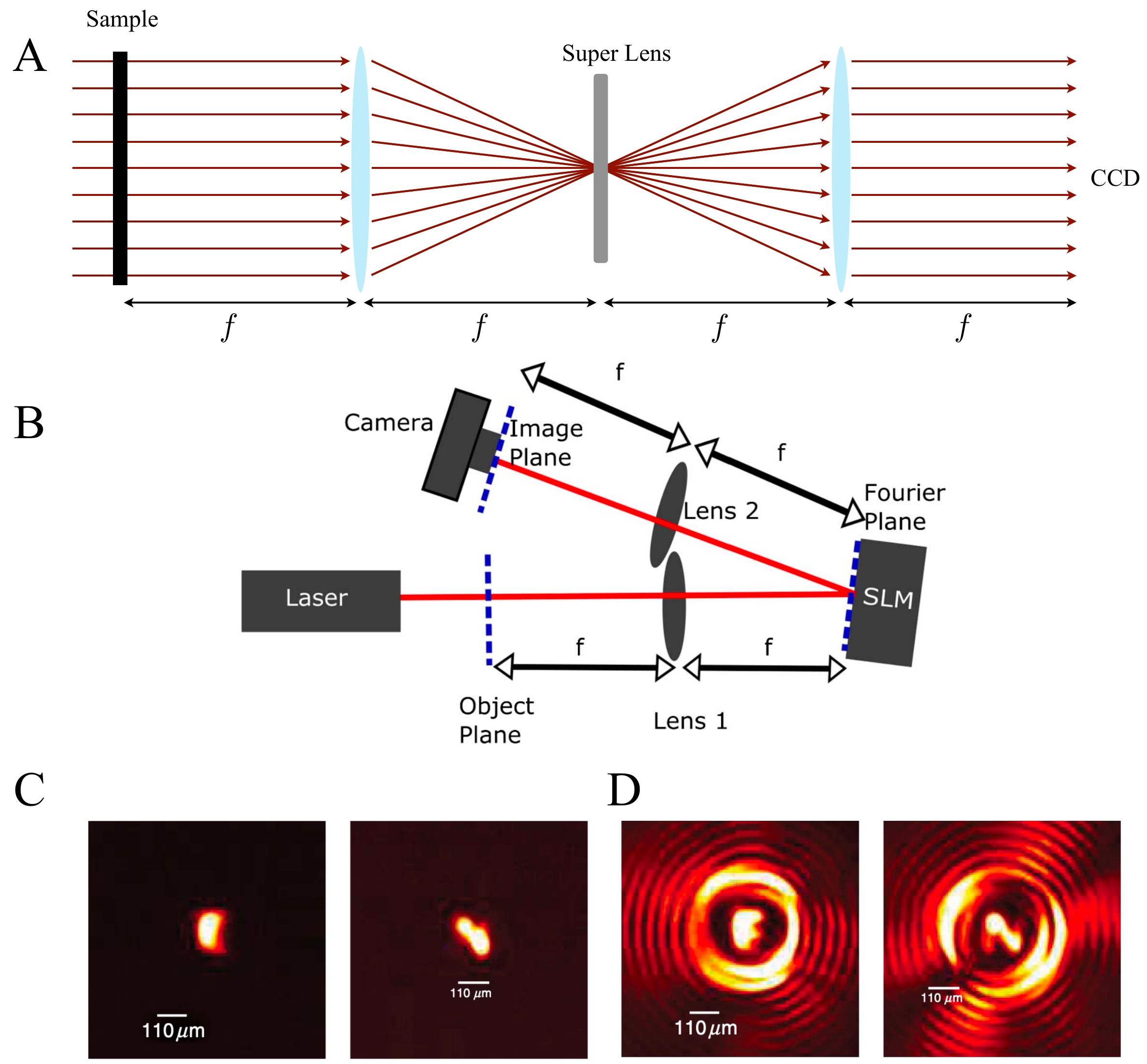}
    \caption{(A) Schematic of a 4f optical imaging system, where the camera and sample are at the same planes, and the super-lens is positioned in the far field of the sample. In the experiment (B), a spatial light modulator (SLM) with an optimised binary hologram serves as the super-lens. A 633-nm HeNe laser illuminates the samples, which consist of Latin letters E (110 $\mu$m$\times$ 87 $\mu$m)) and N (120$\mu$m$\times$ 130$\mu$m). The total imaging system has a numerical aperture (NA) of $0.00864$. (C) Images of both letters captured using a diffraction-limited PSF. (D) Images obtained through superoscillatory PSF microscopy, demonstrating improved resolution. The outer rings in the images result from PSF sidebands located beyond the superoscillation illumination region~\cite{dong:17}.}
    \label{fig:experimentmicroscopy2}
\end{figure}

However, forming an image is more challenging due to the sidelobes, which contain more energy than the hot spot. One approach involved developing a new type of optical beam that forms a ‘needle’ shape, ensuring that the hot spot remains stable while reducing the sideband energies~\cite{rogers:13needle}. Nevertheless, the sideband continues to pose challenges for sub-wavelength imaging for various reasons; for instance, limited field of view and sensitivity of the imaging system are just two significant issues. Several years later, an oscillatory point spread function – a diffraction-limited hotspot surrounded by super-oscillation ripples – was proposed as a solution, utilizing super-oscillatory lenses to post-select the field after illuminating the sample in a $4f$-optical system, rather than generating a hot spot and raster scanning the sample with it~\cite{dong:17}. The super-oscillatory ripples surrounding the point spread function can be engineered to significantly reduce the sidelobe energy, thus enabling integration into an optical imaging system for imaging purposes. Figure~\ref{fig:experimentmicroscopy2} shows the schematic design of this label-free microscopy approach, where the super-oscillatory lens of the point spread function has been employed to capture images of the Latin letters ‘E’ and ‘N’.

When an optical system with a high numerical aperture (NA) is considered, wave propagation becomes non-paraxial, introducing greater complexity to the structure of optical beams at the focus. Richards and Wolf first explored this phenomenon in 1959~\cite{Richards:59}, demonstrating that the (electromagnetic field) intensity distribution at the focal point inherently depends on the pupil’s polarization state. In the early 2000s, Quabis et al.\cite{quabis:00} showed that non-uniform polarization states, particularly those with radial symmetry, produce a tighter focal spot for the electric and a null magnetic field. Conversely, an azimuthally polarized beam generates a strong magnetic field at the focus, and no electric field, as later confirmed experimentally by Dorn et al. a few years later\cite{Dorn:03}.

Further advancements in pupil structuring, such as introducing concentric annular rings with $\pi$-phase jumps, have been explored to achieve a sharper and more longitudinally confined focal spot, beyond the diffraction limit~\cite{wang:08}. Using a radially polarized beam with four annular $\pi$-phase jumps--positioned according to Laguerre polynomials--at a high-NA focus (NA = 1.4) results in a longitudinal intensity profile with a super-oscillatory transverse structure~\cite{kozawa:18}. While this super-oscillatory region is accompanied by sidelobes, increasing the number of phase jumps reduces sidelobe energy, thereby enhancing image resolution. This approach achieves a lateral resolution improvement by a factor of two compared to truncated radially polarized beams.

\subsection{Supergrowth construction}
\label{supergrowth}
In Section \ref{speckle} and \ref{hotspots}, two different mechanisms to create superoscillations and supergrowth, speckle and transmissive hole arrays have been described.  Interestingly, as mentioned, both leverage randomness or pseudo-randomness to create distributions that exhibit superbehavior.  Here we will describe efforts to engineer supergrowing distributions \cite{kr2024experimental}.  Before describing this recent work on supergrowth creation, we point out there has been a large body of work focused on creating superoscillating optical distributions and we direct the reader to a number of excellent reviews (and their references) for details \cite{gbur2019using, chen:19, zheludev:22}.  

\begin{figure}[ht]
\includegraphics[width=0.9\linewidth]{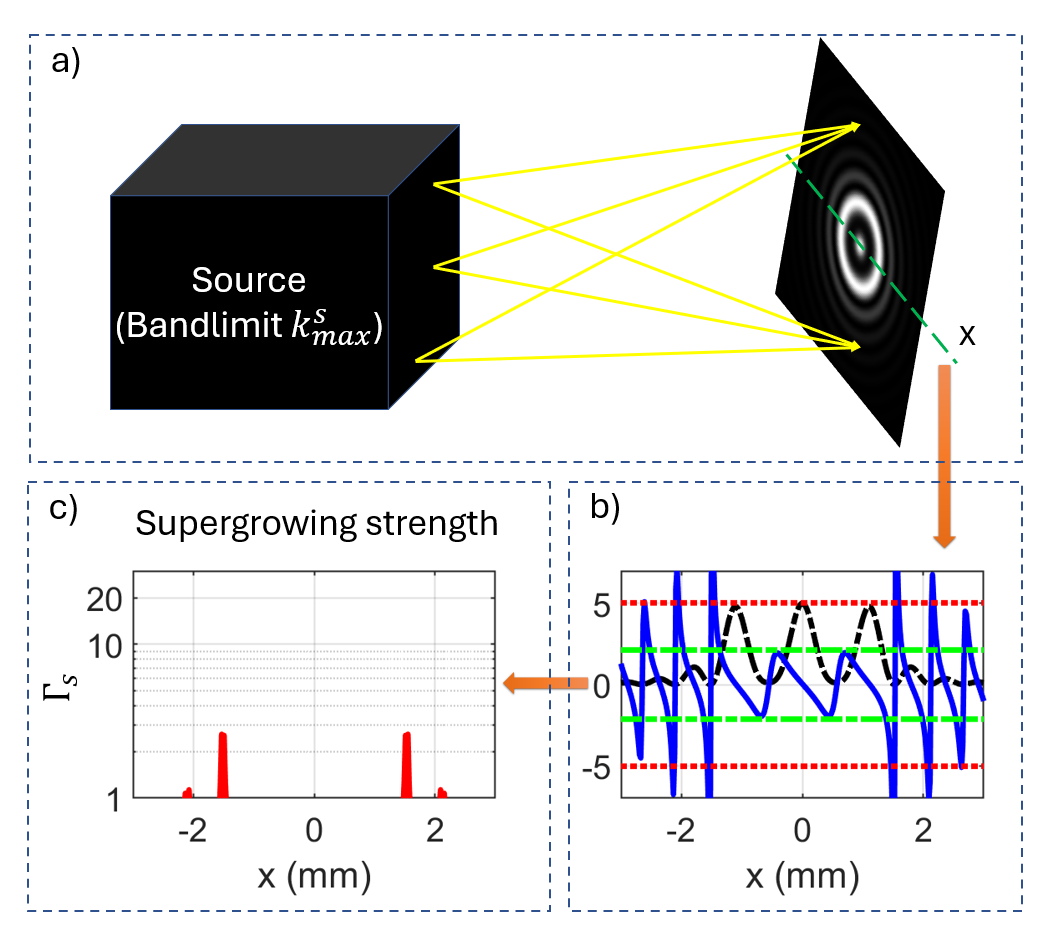}
\caption{\label{SG_fig1} a) The optical system that creates the supergrowing distrbution supports a maximum spatial frequency of $2k^s_{max}$ for irradiance distributions. b) Linecut through one such irradiance distribution.  Illustrated is the irradiance linecut, local growth rate $\kappa$ and the maximum permissible spatial frequency $\pm2k^s_{max}$ (horizontal lines).  c) The locations where the local growth rate exceeds the bandlimit along the line cut in b). Taken from \cite{kr2024experimental}}.
\end{figure}

Figure \ref{SG_fig1}a) is an illustration of the performed experiment.  The optical system that generates the supergrowing irradiance distribution supports a maximum spatial frequency of $k^s_{max}$ ($2k^s_{max}$ for irradiance). The system consists of an intensity stabilized $795$-nm linearly polarized collimated laser beam.  This beam overfills a computer-generated phase-only pixilated hologram (CGPPH) generated using a liquid-crystal-based SLM.  The diffraction from CGPPH is Fourier processed using a classical $4F$ processor consisting of two lenses and a precision pinhole. The intensity corresponding to the complex field at the image plane is measured by a CCD camera.  Figure \ref{SG_fig1}a), on the right side of the panel, presents a simulation of one of the synthesized supergrowing distributions. In Figure \ref{SG_fig1}b), the lower right panel,  a linecut along the simulated irradiance distribution is presented.  In addition to the irradiance, the low growth rate $\kappa$, see Eq. (\ref{localkappa}), and the maximum permissible spatial frequency $\pm2k^s_{max}$ (horizontal lines) are plotted.  In Fig.  \ref{SG_fig1}c), the supergrowing strength $\Gamma(x)=\kappa(x)/2k^s_{max}$ is presented.  This quantifies all locations within the irradiance that surpass the system bandlimit.

\begin{figure}[ht]
\includegraphics[width=0.9 \textwidth]{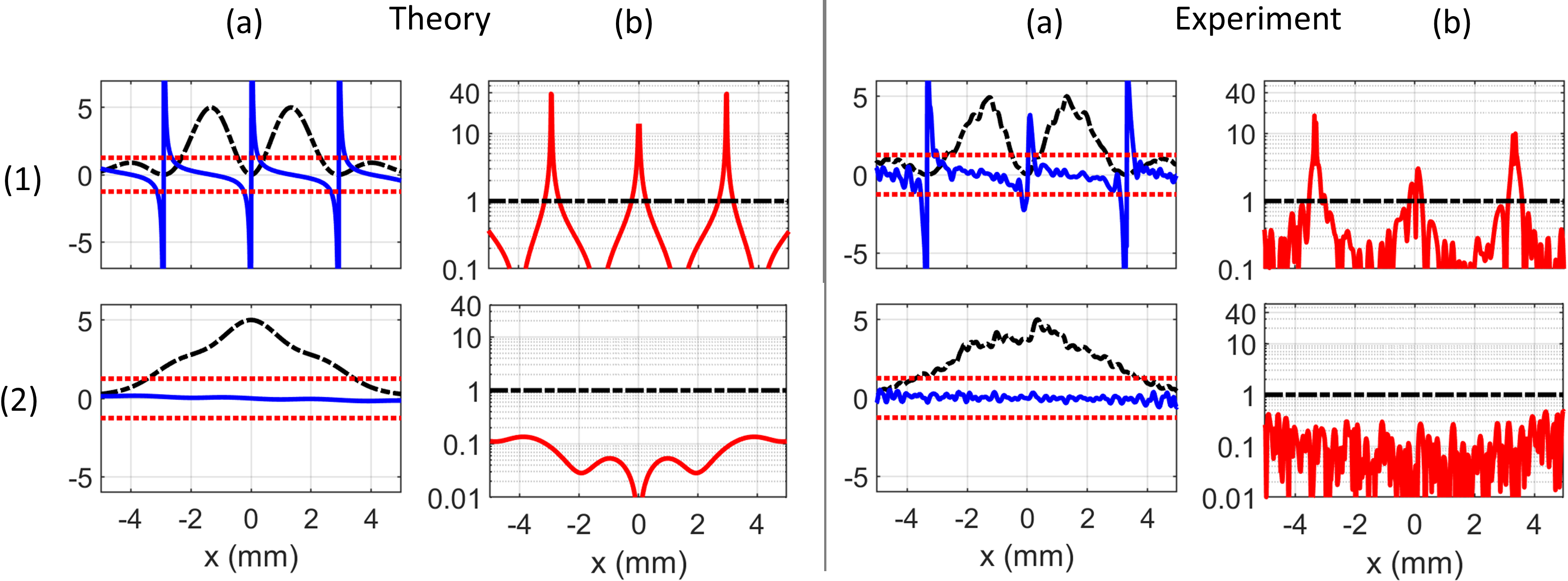}
\caption{\label{SG_fig2} The left four panels present theory and the right four panels are the corresponding experiments.  The left panel column (a) present the irradiance (black dashed line), system bandlimit (horizontal red dashed lines) and the local growth rate $\kappa$ (blue. curve).  Column (b) is the calculated supergrowing strength $\Gamma$.  The right four panels are the corresponding experiments. Taken from \cite{kr2024experimental}.}
\end{figure}

Figure \ref{SG_fig2}a) presents the results of the supergrowth experiments.  The left 4 panels are theoretical calculations and the right 4 panels are experiment.  In the first column of the left panel are the simulated irradiances (black dashed line), the system bandlimit (red dashed horizontal line) and the local growth rate $\kappa(x)$ (blue curve).  The irradiance in the top row is designed to exhibit supergrowth and the bottom row is designed to not have supergrowth.   The second column of the left panel presents the supergrowing strength $\Gamma(x)$.  The right 4 panels present the corresponding experiments.  The theory and experiment demonstrate it is possible to engineer irradiance distributions that possess supergrowth.

\subsection{Recovering superoscillation buried in noise}  \label{recovering}
As discussed earlier, the biggest problem with the practical applications of superoscillations and even supergrowth is simply one of the signal-to-noise ratio being too small.  The superoscillating and even supergrowth signals can be orders of magnitude smaller than the peak amplitude of the signal outside the regions of superbehavior.  This means that even moderate amounts of noise can overwhelm the superoscillating signal.  Recently White \textit{et al.} showed how to recover the superoscillating signal even when it was orders of magnitude smaller than the noise by using a combination of signal construction and filtering \cite{white2024reconstructing}.    

To create robust superoscillations, White \textit{et al.} employed frequency combs of prescribed amplitude and phase.  The idea is to mimic a function $f(x)$ of arbitrary bandwidth over a finite interval using a frequency comb $\psi(x)$ of bandlimited frequency range $\Omega$ given by 
\begin{equation}
    \psi(x) = \sum_{k=0}^{K-1} A_k e^{i \omega_{k} x},
    \label{psi}
\end{equation}
where the amplitudes $A_k$ are complex and $x$ is a set of discretely sampled points. They assumed $N$ equally-spaced frequencies
\begin{equation}
    \omega_k = \frac{k}{K-1} \Omega + \omega_{min},
    \label{omega}
\end{equation} 
ranging from the minimum frequency $\omega_{min}$ to $\omega_{min}+\Omega$ with $\Omega$ being the spectral distance between bandlimits. To solve for the amplitudes, they simply performed matrix inversion
\begin{equation}
    A=M^{+}F,
    \label{F}
\end{equation}
where $M$ is the Moore-Penrose pseudo-inverse form of $e^{i\omega_k x}$ for finite $k$ and $x$, $A$ is the vector of amplitude coefficients , and $F$ is a vector of points on $f$. Since this is a numerical procedure and not an analytic solution (see section 4.4 for the analytic version), the amplitudes $A_k$, will strongly depend on the number of fitted points in $x$ and the interval over which $\psi(x)$ tries to match $f(x)$.  As with all other methods, the bandwidth of $f(x)$ can greatly exceed $\Omega$ leading to superoscillations over the interval.  

By constructing the superoscillating signal in this way, White {\it et al.} created a means by which to perform strong spectral filtering \cite{white2024reconstructing}.  After construction, the signal can be sent to probe a medium of interest.  Under the assumption that the medium is linear and stationary, no additional frequencies will be generated in the medium even though the amplitudes may change.  This means that one simply needs to measure the amplitudes of the spectral components, ignore all other frequencies and then reconstruct the signal through the inverse Fourier transform.  By doing so, the bulk of the noise can be rejected.    

\begin{figure}[ht]
\begin{center}
\includegraphics[width=.7\textwidth]{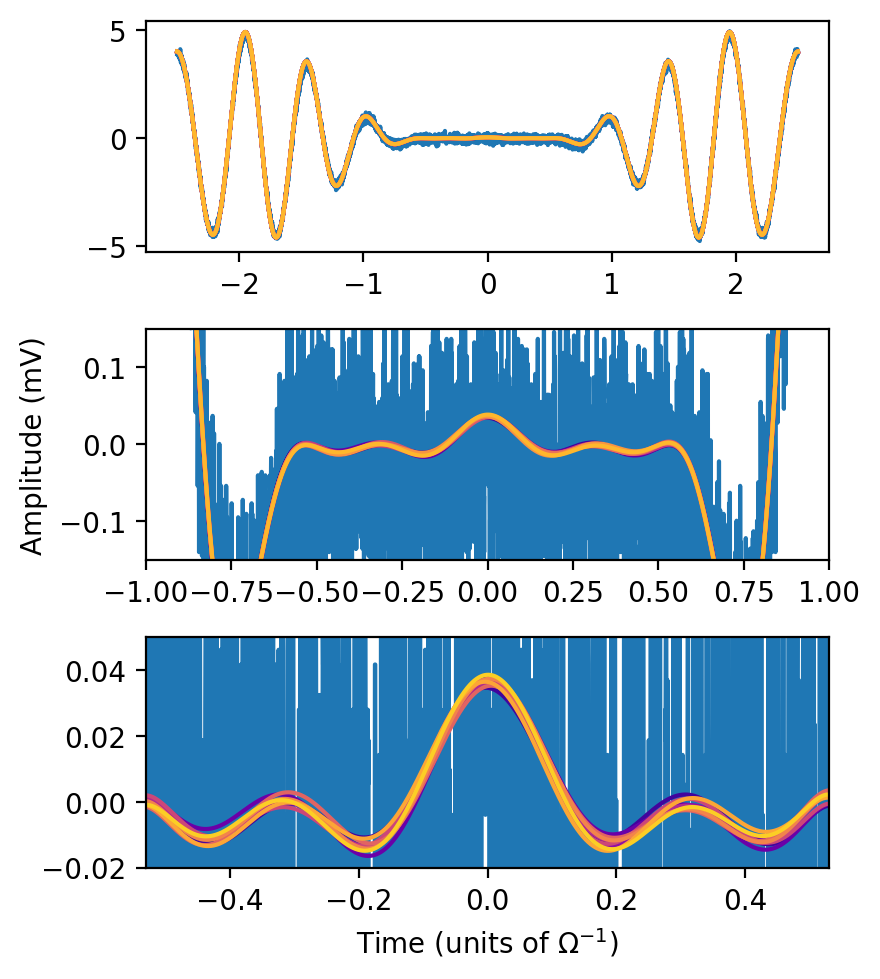}
\end{center}
    \caption{A superoscillatory sinc function with four times the local bandwidth of the entire signal is shown. The noise amplitude is significantly larger than the superoscillation.  The top figure shows a complete cycle of the periodic waveform.  The middle and bottom figures show the signal with noise (blue) and several reconstructions of the signal after filtering. After Ref.~\cite{white2024reconstructing}.} 
    \label{Reconstruct}
\end{figure}

The noise rejection can be easily understood in terms of Parseval's theorem.  For a periodic function and white Gaussian noise, the periodic function will increase its energy with more samples while the noise will remain, on average, constant in the frequency space.

To demonstrate this method's robustness to noise White \textit{et al} demonstrated a proof of concept in a guided wave experiment with results shown in Fig. \ref{Reconstruct}.  A superoscillating signal consisting of a sinc function with a local bandwidth four times larger than $\Omega$. White noise with an amplitude 17 dB larger than the superoscillation amplitude was added to the signal.  Using the filtering technique, a high-fidelity filtered version of superoscillation signal was recovered. With 10 averages of the signal, they were able to get a mean squared error of about 1\% in the reconstructed signal.

\section{Methods of Constructing Super Point Spread Functions}  \label{methods}

In the early years of superoscillation research, the functions that exhibited superoscillations were discovered on a case by case basis - most famously Eq.~(\ref{f}).  However, it gradually became clear that a more systematic way to understand and generate this class of functions was needed.  In this section, we cover a few of the most influential ways to generate superoscillating functions.

\subsection{Method of Forced Zeros} \label{Gbur}
\begin{figure}
    \centering
    \includegraphics[width=0.5\linewidth]{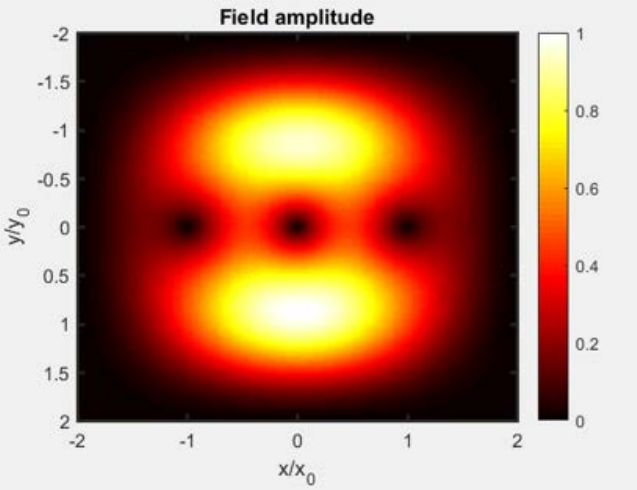}
    \caption{After Ref.~\cite{smith2020mathematical}.
    Illustration of the method of Smith and Gbur to construction superoscillatory functions.  The axes are plotting in units of inverse bandwidth, $x_0, y_0 = 1/\Omega$.  Zeros in the real space are inserted at $x/x_0 = (-1, 0, 1)$.  The plot uses the values $n=m=6$.}
    \label{fig:Gbur}
\end{figure}

In this subsection, we review the method described by Smith and Gbur to create superoscillatory functions \cite{smith2020mathematical}.  They focus on two dimensional function appropriate for the imaging plane of an optical system.  Their method is to specify the band limit $\Omega$, and then have a multistep process:
\begin{itemize}
\item 
Impose the bandlimit with a power of a trigometric function in the Fourier space
\be
{\tilde f}(k_x, k_y) = \cos\left( \frac{\pi k_x}{\Omega}\right)^n \cos\left( \frac{\pi k_y }{ \Omega}\right)^m,
\ee
where the values of the wavenumbers is limited to $-\Omega/2 \le k_x , k_y \le \Omega/2$, so that the cosine function makes the function go to 0 at the band limits.  The values of $n,m$ are arbitrary.

\item
The function is Fourier transformed to the real space, $f(x, y)$, and then multiplied by zeros at desired locations:
\be
g(x, y) = f(x, y) \prod_j(x-x_j)(y-y_j).
\ee
The multiplication of the function by linear function does not change the bandlimit.  This can easily be seen because the product of two functions in the real space leads to a convolution of the Fourier transforms in the Fourier space.  The Fourier transform of any polynomial only has weight at zero frequency.  There the convolution cannot exceed the bandlimit of the original function.
\end{itemize}
The placement of zeros within the diffraction limited space then forces the function to oscillate faster than the bandlimit as the function goes through the prescribed zeros.  This is similar in spirit to the method of Kempf; see his first aspect \cite{kempf2018four}.  The method is illustrated to produce the superoscillating function in Fig.~\ref{fig:Gbur}.

\subsection{Canvas Function Method}  \label{canvas}
The next superoscillation technique we discuss was developed by \v{S}oda and Kempf \cite{vsoda2020efficient}.  Their technique relies on two main ideas.  First, all orders of sinc functions of the type:
\begin{equation}
    c_m(x)=\textrm{sinc}\left(\frac{\Omega x}{m}\right)^m,
\end{equation}
share the same bandlimit $\Omega$.  Here, $c_m$ is the $m^{th}$ order canvas function.  The name ``canvas'' comes from the idea that it represents a blank slate upon which another function can be written. Importantly, the numerator of the canvas function is periodic and the denominator is a polynomial of power $m$.  This behavior is needed for square integrability.  

The second feature is that any function over an interval can be approximated by an $n^{th}$-order power series: 
\begin{equation}
    f_n(x)=\sum_{k=0}^{n-1}a_k x^k.
\end{equation}
The superoscillating function is then given by the product:
\begin{equation}
    g(x):=f_n(x)c_m(x).  
\end{equation} 
For $m>n+1$ this is a square-integrable function. While this function methodology works in general, it has also been applied to superoscillations, because an arbitrary function, without bandwidth constraints, can be approximated over the interval.  Unfortunately, if the local bandwidth of the function exceeds the bandwidth of the canvas functions, the all-too-familiar large lobes outside the interval are also generated.       

\subsection{Finite Order Taylor Expansion Method} \label{Taylormethod}

In the work of Aharonov {\it et al.}, another method is used to construct superoscillatory functions.  The authors consider a generalized Fourier series,
\be
f_N = \sum_{j=0}^N X_j(N, a) e^{i k_j(N) x}, 
\ee
where one choice for the Fourier wavenumbers that can be made is $k_j(N) = (1- 2 j/N)$.  The bandlimit is 1.  The goal is to find in the mathematical limit $N \rightarrow \infty$ that $f_N$ converges to $f$, a function that is not bandlimited, and superoscillates in general.  The strategy is to match the Taylor expansion of these two functions up to order $N$, such that
\be
f^{(k)}(x=0) = f_N^{(k)}(x=0),
\ee
so the order of derivative runs from 0 to $N$. 
There are $N$ coefficients $X_j$ in this construction, so taking $N$ derivatives of the target function and taking their value at the origin gives a set of $N$ equations that allow a solution that is neither under nor over determined.  While other possibilities can be conceived, this is a simple approach to find the bandlimited coefficients.

The specific example the authors consider is $f = e^{i a x}$, where $a>1$ corresponds to a superoscillating function, that is clearly outside the band limit.  Making the constructive solution for each value of $N$, as $N \rightarrow \infty$ the range of superoscillation of $f_N(x)$ extends over all space as $f_N$ converges to $f$, while remaining bandlimited for every finite value of $N$.  The authors find an explicit expression for finite $N$
\be
X_j(N, a) = \prod_{i=0, k \ne j}^N \left(\frac{k_i(N) - a}{k_i(N) - k_j(N)}\right)
\ee
using the mathematics of Vandermonde matricies.
The method extends to more general expression of the wavenumber $k_i(N)$ as well as other superoscillating functions, but the program must be numerically implemented in general.
\subsection{Bandlimited Approximation Theory on an Interval} \label{approximationtheory}

Karmakar and Jordan went beyond the concept of superoscillation and considered the following more general problem \cite{karmakar2023beyond}:  Given a function that is not bandlimited, what is the best approximation to this function in a finite interval, that is bandlimited?  They answered this problem by first making a basis of bandlimited functions.  We review their construction both for a finite interval $(x_1, x_2)$, and then for the whole real line.

\begin{figure}
    \centering
    \includegraphics[width=0.8\linewidth]{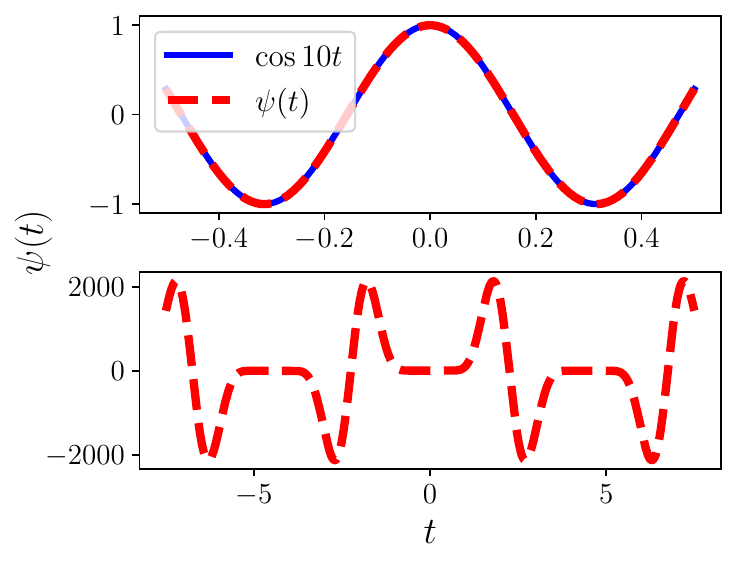}
    \caption{After Ref.~\cite{karmakar2023beyond}.
The top panel shows the target function ${\rm cos}(10 t)$ as a solid blue curve, and the approximating function $\psi(t)$ in dashed red.  The interval of interest is $t \in [-1/2, 1/2]$.  The bottom panel shows the function on a larger interval, illustrating the large function values needed to achieve the approximation beyond the band limit of the function, which is $2\pi$. }
    \label{fig:anjmethod}
\end{figure}

For a finite interval, consider the Fourier series corresponding to a bandlimited function
\be
\psi_N(x) = \sum_{n=0}^N C_n e^{i k_n x}, \label{psi2}
\ee
where $k_n = 2\pi (1-2n/N)$ and $N$ is a positive integer.  The band limit is here $2\pi$.  Now consider a function that is not contained in this set of bandlimited functions, $\Phi(x)$.  The goal is to mimic this function as closely as possible in the finite interval $(x_1, x_2)$.  To do so, define an error function
\be
\epsilon(x_1, x_2)  = \int dx |\Phi(x)- \psi_N(x)|^2dx, \label{epsilon2}
\ee
that captures any deviation between the target function and the bandlimited function on the interval of interest.  Ref.~\cite{karmakar2023beyond} shows that by minimizing this function as a variation on the coefficients $C_n$ leads to a matrix equation,
\be
{\bf C} = {\bf \alpha}^+ {\bf b},
\ee
where ${\bf C}$ is the vector of unknown variational coefficients $C_n$, the vector $\bf b$ is defined via it components
\be
b_n = \int_{x_1}^{x_2} dx \Phi(x) e^{-i k_n x} dx.
\ee
The matrix ${\bf \alpha}^+ $ is the Moore-Penrose pseudo-inverse of the matrix $\alpha$, whose elements are
\be
\alpha_{nm} =  \int_{x_1}^{x_2} dx e^{i(k_m - k_n)x}.
\ee
This gives a constructive solution to the coefficients $C_n$ that minimizes the error on the interval.  Such a solution can be used to find any desired superoscillatory function, but is more general than that:  Any functional dependence that eludes bandlimited functions can be approximated, even functions that have discontinuities.  In practice, a finite dimension of the system is chosen, and the matrices and vectors are calculated and manipulated numerically.  The method is illustrated in Fig.~\ref{fig:anjmethod}, where the function $\cos(10 t)$ is well approximated on the interval $(-1/2, 1/2)$ using $N=9$ terms in the Fourier series.

For the whole real line, it is essential to move away from Fourier series to Fourier transforms.
Given any such function $f(x)$, its Fourier transform ${\tilde f}(k)$ is chosen without loss of generality to have compact support for $-1 \le k \le 1$.  This Fourier transform is then expanded in a complete basis of orthogonal polynomials,
\be
{\tilde f}(k) = \sum_{n=0}^\infty d_n P_n(k),
\ee
where $d_n$ are in general complex arbitrary coefficients, and the choice of $P_n$ to be the Legendre polynomials was made, forming a natural basis on (-1,1).  Using the orthogonality property of the Legendre polynomials,
\be
\int_{-1}^1 dk P_n(k) P_m(k) = \frac{2}{2n+1} \delta_{nm}, 
\ee
the coefficients $d_n$ are then defined as
\be
d_n = \frac{2n+1}{2} \int_{-1}^1 dk {\tilde f}(k) P_n(k).
\ee
In the original space, the Fourier transform of Legendre polynomials are spherical Bessel functions $j_n(x)$,
\be
f(x) = \sqrt{\frac{2}{\pi}} \sum_{n=0}^\infty i^n d_n j_n(x).
\ee
The spherical Bessel functions are individually band-limited functions, and can be constructed from derivatives of the sinc function. 
\be
j_n(x) = (-x)^{n}\left( x^{-1} \frac{d}{dx}\right)^n {\rm sinc}(x). 
\ee
Thus, this choice of basis is natural for band-limited functions.

A similar procedure as described for the finite interval can be applied to this example, where the authors truncate the series of spherical Bessel functions at a finite order $N$, to get the function $f_N$ and the goal is to mimic a target function $g(x)$ on the finite interval $(x_1, x_2)$.
By minimizing the same error function (\ref{epsilon2}) with the function $f_N(x)$ and $g$, we arrive at a similar matrix equation
\be
{\bf D} = {\bf A}^+ {\bf B},
\ee
where $\bf D$ is a vector of coefficients $D_n$ the vector $\bf B$ has coefficients
\be
B_n = \sqrt{\pi/2} \int_{x_1}^{x_2} dx g(x) 
j_n(x),
\ee
and the matrix $ {\bf A}^+$ is the pseudo-inverse of the matrix $A$ defined via its elements
\be
A_{nm} = \int_{x_1}^{x_2} dx j_n(x) j_m(x).
\ee
This completes the solution.


\subsection{Technology: Spatial light modulators, arbitrary waveform generators}  \label{technology}
Superoscillatory wavepackets require precise engineering to tailor their internal degrees of freedom to match functions exhibiting superoscillatory characteristics, such as the one in Eq.~(1), as discussed in the previous sections of this chapter. Achieving this demands full control over the amplitude and phase of the wavefunction, which is a challenging task, as elaborated in the following section. Without loss of generality, we will focus on the optical domain, though the approach is also applicable to massive particle wavepackets, except for spin state control.

In the optical domain, where electromagnetic waves are considered, the beam’s amplitude, phase, and polarization must be carefully engineered to achieve the desired super-oscillatory behavior. Complete control over these three degrees of freedom (amplitude, phase, and polarization) is essential when working in the non-paraxial regime, where the beam size $w_0$ is comparable to the wavelength $\lambda$. Optical beams generated in the laboratory predominantly fall within the paraxial wave equation regime unless an optical system with a numerical aperture greater than 0.7 is employed, typically under a tight-focusing condition~\cite{Richard:59}.

This implies that precise control over the beam’s amplitude and phase at a specific transverse plane is necessary. Once the beam is generated at a given plane, it can evolve to other planes through (unitary) free-space propagation. This approach simplifies the problem by shaping the beam's phase and amplitude within the paraxial regime at a given transverse plane. The solution to this problem is unique, meaning that a properly designed initial wavepacket will deterministically evolve into a well-defined field upon free-space propagation (or propagation through a known linear medium). Thus, by carefully engineering the initial wavepacket at a given plane, one can reliably generate super-oscillatory wavepackets with the desired properties. Of course, extending beyond the paraxial regime requires spatially varying polarization, amplitude, and phase engineering, which can be achieved through proper \emph{inverse engineering}, either numerically~\cite{Wu:18} or, in some specific cases, analytically. However, in this discussion, we focus on the paraxial regime and the techniques for shaping electromagnetic fields into a specific wavefunction $\psi(x,y,z)$. This technique is also applicable to matter waves, such as electrons and neutrons.

Consider a paraxial optical beam with amplitude $A(x,y,z)$  and phase $\chi(x,y,z)$, expressed as  $\psi(x,y,z) = A(x,y,z) e^{i\chi(x,y,z)}$  in Cartesian coordinates $(x,y,z)$~\cite{siegman:86}. This beam satisfies the paraxial wave equation (PWE):
\begin{eqnarray}\label{eq:PWE}
\left(\partial^2_x + \partial^2_y - 4ik\partial_z\right) \psi(x,y,z) = 0,
\end{eqnarray}
where  $\partial_i = \frac{\partial}{\partial i}$  represents the partial derivative with respect to coordinate  $i$ .

At any given plane $z$, the beam can be expressed as a convolution of the field at the initial plane  $(x_0, y_0, 0)$ with the propagation kernel  $K(x,y,z; x_0, y_0, 0)$, which can be obtained via the PWE. This is known as the Fresnel propagator in optics, though the same concept applies to matter waves. Consequently, the initial wavefunction
\begin{eqnarray}
\psi(x_0, y_0, 0) = A(x_0, y_0, 0) e^{i\chi(x_0, y_0, 0)},
\end{eqnarray}
must be precisely engineered.

There are two primary approaches to achieve this: one involves using separate devices to impose the amplitude  $A(x_0, y_0, 0)$  and phase  $e^{i\chi(x_0, y_0, 0)}$  independently at a given plane, while the other relies on a single device--namely, a phase-only modulator--to shape both simultaneously. The first approach, though conceptually straightforward, is challenging and expensive because it requires two optical elements (one for amplitude and one for phase) that must be perfectly aligned. The second approach, using a phase-only hologram (kinoform), simplifies alignment but is less efficient, as most of the incoming optical power is not converted into the desired amplitude distribution due to the inherent limitations of phase-only modulation. Various methods exist for generating optical beams using phase-only holograms (see, e.g. Ref.~\cite{Arrizon:07}). However, many of these techniques suffer from low beam generation fidelity or difficulties in handling beams whose amplitude varies as a function of  $x$  and $y$. To address these limitations, we focus on the approach developed in 2013~\cite{Bolduc:13}, which offers improved field generation fidelity. This method is now widely used to generate arbitrary optical beams with spatial light modulators (SLMs). Subsequently, alternative approaches were proposed that specifically aimed at generating super-oscillatory functions, e.g., see Ref.~\cite{zhu2014arbitrary}.

SLMs are modern liquid-crystal-based optical devices commonly employed to shape optical fields in the UV, visible, and NIR domains. By applying voltage to individual pixels--typically around  $8 \mu\text{m} \times 8 \mu\text{m}$  or slightly larger--the orientation of liquid crystal molecules is altered, causing variations in the optical path length experienced by the passing light. As a result, SLMs introduce controlled wavefront (phase) modifications to an incoming beam, represented as $e^{i\Phi(x,y)}$. SLMs can be either transmissive or reflective. Combined with a polarizer, they can function as amplitude modulators, with commercially available options supporting both functionalities, i.e. either phase or amplitude masking. However, it is important to note that the $x$ and $y$ coordinates in these devices are discrete rather than continuous variables, which imposes certain constraints on beam-shaping precision. If we assume a uniform, flat-phase incoming beam, the outgoing beam from the spatial light modulator (SLM) is given by $e^{i\Phi(x,y)}$, representing only phase modulation. However, two challenges arise: (1) The phase modulation $\Phi(x,y)$ is not continuous and does not precisely span the full  $2\pi$  range with arbitrary resolution. Instead, it is quantized, typically in 8-bit steps (0-255), and depends on the gamma function of the modulation; and (2) How can we modify the beam’s amplitude using only-phase modulation?\newline

The first challenge can be addressed through a holographic approach, generating a computer-generated hologram (CGH) via numerical simulation, where the desired phase is interfered with a uniform but tilted beam. This process results in a hologram that possesses a diffraction grating with a specific pitch size $\Lambda_x$, determined by the diffraction angle in the CGH. When the SLM is illuminated with a uniform plane wave, the diffracted beam acquires a specific phase pattern encoded in the hologram.

There are several standard techniques for generating computer-generated holograms with adjustable diffraction efficiency. The most common methods include~\cite{rosi2022theoretical}:
\begin{itemize}
\item{\emph{Binary grating:}} $\pi (1 + \text{Sign}(\sin{(\Phi(x_0,y_0,0) + 2\pi/\Lambda_x)}))$
\item{\emph{Sinusoidal grating:}}  $\pi(1 + \sin{(\Phi(x_0,y_0,0) + 2\pi/\Lambda_x)})/2 $
\item{\emph{Blazed grating:}}  $\text{Mod}((\Phi(x_0,y_0,0) + 2\pi/\Lambda_x), 2\pi)$,
\end{itemize}
where Sign is the signature function taking values of $\pm1$, Mod is the modulo function, and $\Lambda_x$ is the pitch of the grating.

For our purposes, the blazed grating--a hologram designed with a modulus $2\pi$ function--is of particular interest, as it maximizes diffraction efficiency at the first-order diffraction. If the modulation is perfect, then according to Fourier expansion, blazed function modulation theoretically achieves 100\% efficiency at the first diffraction order. To address the second challenge--introducing specific amplitude control to the diffracted beam while maintaining the same phase pattern using a phase-only SLM--the modulation function must be adjusted according to the amplitude $A(x_0,y_0,0)$. 
The hologram is thus given by:
\begin{eqnarray}\label{eq:hologrammodulation}
M(x_0,y_0,0) \, \text{Mod}((\Phi(x_0,y_0,0) + 2\pi/\Lambda_x), 2\pi).
\end{eqnarray}
After performing a Taylor-Fourier expansion, the diffracted beam at the first-order diffraction is given by:
\begin{eqnarray}\label{eq:filed1stdiffraction}
-\text{sinc}(\pi M(x_0,y_0,0)-\pi) \, e^{i(\Phi(x_0,y_0,0) + \pi M(x_0,y_0,0))},
\end{eqnarray}
where $\text{sinc}(.)=\frac{\sin(.)}{(.)}$ is the sinc function. This expression shows how phase modulation can effectively control the amplitude of the diffracted beam via the Sinc function. The modulation depth $M(x_0,y_0,0)$ directly influences the beam’s amplitude, making it possible to encode both amplitude and phase information simultaneously using a single phase-only SLM.

To generate the desired beam $A(x_0,y_0,0) e^{i\chi(x_0,y_0,0)}$, we apply \emph{inverse engineering}, yielding the following relations:
\begin{eqnarray}
M(x_0,y_0,0) = 1+\frac{1}{\pi}\text{sinc}^{-1}(A(x_0,y_0,0)),\\ \nonumber
\Phi(x_0,y_0,0)=\chi(x_0,y_0,0)-\pi M(x_0,y_0,0).
\end{eqnarray}
Here, $\text{sinc}^{-1}(.)$ denotes the inverse sinc function within the domain $[-\pi,0]$. This approach allows precise control over both the amplitude and phase of the diffracted beam using a phase-only spatial light modulator.
\begin{figure}
    \centering
    \includegraphics[width=1\linewidth]{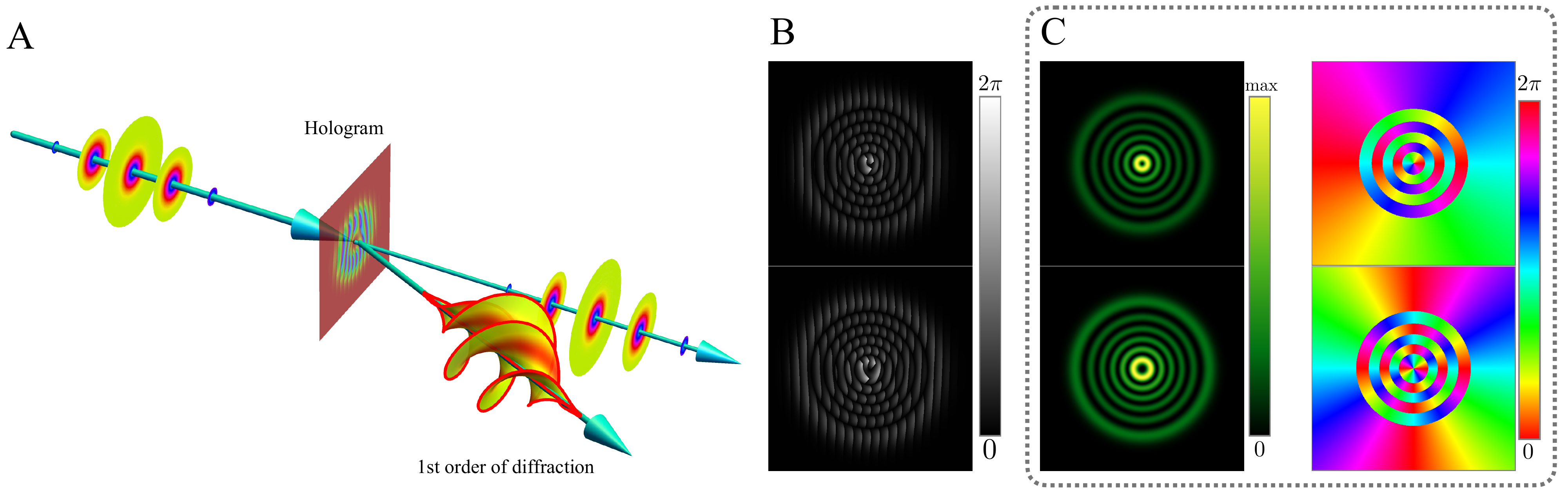}
    \caption{After Ref.~\cite{kozawa:18}.  (A) Schematic illustration of the generation of a desired optical beam at the first-order diffraction when a plane wave illuminates an amplitude-phase hologram. (B) Amplitude-phase holograms designed to generate Laguerre-Gaussian (LG) modes with radial index $p = 5$ and azimuthal indices $m = 1$ and $m = 2$. (C) The intensity and phase profiles of the corresponding LG beams generated from the holograms are shown in (B).}
    \label{fig:slm}
\end{figure}

Figure~\ref{fig:slm} shows a schematic illustrating how a desired optical beam can be generated using a single phase-only spatial light modulator (SLM). In (B) examples of amplitude-phase holograms designed to produce Laguerre-Gaussian (LG) modes with radial and azimuthal indices $p=5, m=1$ and $p=5, m=2$ are also shown. Such holograms have recently been utilized to generate optical beams with superoscillatory characteristics, which are increasingly being applied in advanced imaging techniques~\cite{kozawa:18}.

\section{Conclusions and Outlook} 
\label{conclusions}
In this selective review, we have discussed the phenomena of superoscillations and how it can be applied to various applications in different physical systems. We have written here on the domain of most activity: electromagnetic waves, focusing on the optical domain. We discussed applications ranging from superresolution to obtain optical images with a focus better than that allowed by the Raleigh criterion to resolving spectroscopic signatures of molecules better than allowed by the signal's bandwidth.  Various methods to produce optical superoscillations were reviewed, including quasi-periodic hole arrays producing hot spots, spatial light modulator methods to engineer point spread functions, and even the simple propogation through random media to produce superoscillatory and supergrowing speckle patterns.  There are now many different mathematical techniques to produce superoscillatory and supergrowing functions, and even systematic ways to approximate functions that lay outside the space of bandlimited functions in an interval that were also discussed.

While this review already covers much material, there are many topics we did not touch.  Perhaps the most glaring is the subject from which superoscillations emerged from in the first place:  quantum mechanics and the theory of measurement \cite{jordan2024quantum}.  Indeed, this subject originated in the investigation of matter waves \cite{aharonov1847can}.  Recent work has generalized the concept of the superoscillation to a superobservable \cite{jordan2024superphenomena}.  In that work, a reconnection to the weak value \cite{PhysRevLett.60.1351} showed that the local wavenumber exceeding the bandlimit of a signal is identical to the weak value of the momentum operator exceeding the eigenvalue range of that operator when the pre-selected wavefunction is of the superoscillatory type, and the postselection is the position.  By replacing the momentum operator by a general operator, a suitable criterion for superbehavior could be defined.  In addition to quantum waves, superoscillators phenomenona and its relative can be found in any wave medium:  Acoustic waves, water waves, and so on. A recent outgrowth of superoscillation is achieving super range resolution in radar.  The methods introduced in ``superradar'' use tailored waveforms combined with parameter estimation in order to achieve discrimination of multiple targets orders of magnitude beyond traditional radar theory permits \cite{howell2023super,PhysRevResearch.6.033341,PhysRevApplied.20.064046}.  The fact that many new concepts and applications in this area continue to emerge points to a high research activity outlook in this area for the foreseeable future.


\begin{thebibliography}{10}
\providecommand{\url}[1]{\texttt{#1}}
\providecommand{\urlprefix}{URL }
\providecommand{\doi}[1]{https://doi.org/#1}

\bibitem{aharonov2011some}
Aharonov, Y., Colombo, F., Sabadini, I., Struppa, D., Tollaksen, J.: Some mathematical properties of superoscillations. Journal of Physics A: Mathematical and Theoretical  \textbf{44}(36),  365304 (2011)

\bibitem{aharonov1847can}
Aharonov, Y., Popescu, S., Rohrlich, D.: How can an infra-red photon behave as a gamma ray. Tel-Aviv University Preprint TAUP  \textbf{90}, ~1990 (1847)

\bibitem{PhysRevLett.60.1351}
Aharonov, Y., Albert, D.Z., Vaidman, L.: How the result of a measurement of a component of the spin of a spin-1/2 particle can turn out to be 100. Phys. Rev. Lett.  \textbf{60},  1351--1354 (1988)

\bibitem{Arrizon:07}
Arriz\'{o}n, V., Ruiz, U., Carrada, R., Gonz\'{a}lez, L.A.: Pixelated phase computer holograms for the accurate encoding of scalar complex fields. J. Opt. Soc. Am. A  \textbf{24}(11),  3500--3507 (Nov 2007)

\bibitem{baranova1981dislocations}
Baranova, N., Zel'Dovich, B.Y., Mamaev, A., Pilipetskiǐ, N., Shkukov, V.: Dislocations of the wavefront of a speckle-inhomogeneous field (theory and experiment). ZhETF Pisma Redaktsiiu  \textbf{33}, ~206 (1981)

\bibitem{berry1994evanescent}
Berry, M.V.: Evanescent and real waves in quantum billiards and gaussian beams. Journal of Physics A: Mathematical and General  \textbf{27}(11), ~L391 (1994)

\bibitem{berry1978disruption}
Berry, M.: Disruption of wavefronts: statistics of dislocations in incoherent gaussian random waves. Journal of Physics A: Mathematical and General  \textbf{11}(1), ~27 (1978)

\bibitem{berry2014superoscillations}
Berry, M.: Superoscillations, endfire and supergain. In: Quantum Theory: A Two-Time Success Story: Yakir Aharonov Festschrift, pp. 327--336. Springer (2014)

\bibitem{berry2000phase}
Berry, M., Dennis, M.: Phase singularities in isotropic random waves. Proceedings of the Royal Society of London. Series A: Mathematical, Physical and Engineering Sciences  \textbf{456}(2001),  2059--2079 (2000)

\bibitem{berry2008natural}
Berry, M., Dennis, M.: Natural superoscillations in monochromatic waves in d dimensions. Journal of Physics A: Mathematical and Theoretical  \textbf{42}(2),  022003 (2008)

\bibitem{berry2006evolution}
Berry, M., Popescu, S.: Evolution of quantum superoscillations and optical superresolution without evanescent waves. Journal of Physics A: Mathematical and General  \textbf{39}(22), ~6965 (2006)

\bibitem{berry2019geometry}
Berry, M., Shukla, P.: Geometry of 3d monochromatic light: local wavevectors, phases, curl forces, and superoscillations. Journal of Optics  \textbf{21}(6),  064002 (2019)

\bibitem{Bolduc:13}
Bolduc, E., Bent, N., Santamato, E., Karimi, E., Boyd, R.W.: Exact solution to simultaneous intensity and phase encryption with a single phase-only hologram. Optics letters  \textbf{38}(18),  3546--3549 (2013)

\bibitem{chen:19}
Chen, G., Wen, Z.Q., Qiu, C.W.: Superoscillation: from physics to optical applications. Light: Science \& Applications  \textbf{8}(1), ~56 (2019)

\bibitem{cox1986practical}
Cox, H., Zeskind, R., Kooij, T.: Practical supergain. IEEE Transactions on Acoustics, Speech, and Signal Processing  \textbf{34}(3),  393--398 (1986)

\bibitem{dennis2008superoscillation}
Dennis, M.R., Hamilton, A.C., Courtial, J.: Superoscillation in speckle patterns. Optics letters  \textbf{33}(24),  2976--2978 (2008)

\bibitem{dong:17}
Dong, X.H., Wong, A.M., Kim, M., Eleftheriades, G.V.: Superresolution far-field imaging of complex objects using reduced superoscillating ripples. Optica  \textbf{4}(9),  1126--1133 (2017)

\bibitem{Dorn:03}
Dorn, R., Quabis, S., Leuchs, G.: Sharper focus for a radially polarized light beam. Phys. Rev. Lett.  \textbf{91},  233901 (Dec 2003). \doi{10.1103/PhysRevLett.91.233901}, \url{https://link.aps.org/doi/10.1103/PhysRevLett.91.233901}

\bibitem{dressel2014colloquium}
Dressel, J., Malik, M., Miatto, F.M., Jordan, A.N., Boyd, R.W.: Colloquium: Understanding quantum weak values: Basics and applications. Reviews of Modern Physics  \textbf{86}(1),  307--316 (2014)

\bibitem{quabis:00}
Eberler, M., Gl{\"o}ckl, O.: Focusing light to a tighter spot. Optics communications  \textbf{179}(1-6), ~1--7 (2000)

\bibitem{gbur2019using}
Gbur, G.: Using superoscillations for superresolved imaging and subwavelength focusing. Nanophotonics  \textbf{8}(2),  205--225 (2019)

\bibitem{goodman:05}
Goodman, J.W.: Introduction to Fourier optics. Roberts and Company publishers (2005)

\bibitem{goodman2007speckle}
Goodman, J.W.: Speckle phenomena in optics: theory and applications. Roberts and Company Publishers (2007)

\bibitem{hastings2025localization}
Hastings, R.L., Webb, K.J.: Localization and coherent imaging of hidden moving objects using laser speckle. Optics Letters  \textbf{50}(4),  1172--1175 (2025)

\bibitem{haviland1995supergain}
Haviland, R.: Supergain antennas: possibilities and problems. IEEE Antennas and Propagation Magazine  \textbf{37}(4),  13--26 (1995)

\bibitem{heeman2019clinical}
Heeman, W., Steenbergen, W., van Dam, G.M., Boerma, E.C.: Clinical applications of laser speckle contrast imaging: a review. Journal of biomedical optics  \textbf{24}(8),  080901--080901 (2019)

\bibitem{STED:94}
Hell, S.W., Wichmann, J.: Breaking the diffraction resolution limit by stimulated emission: stimulated-emission-depletion fluorescence microscopy. Optics letters  \textbf{19}(11),  780--782 (1994)

\bibitem{howell2023super}
Howell, J.C., Jordan, A.N., {\v{S}}oda, B., Kempf, A.: Super interferometric range resolution. Physical Review Letters  \textbf{131}(5),  053803 (2023)

\bibitem{huang2007optical}
Huang, F.M., Chen, Y., De~Abajo, F.J.G., Zheludev, N.I.: Optical super-resolution through super-oscillations. Journal of Optics A: Pure and Applied Optics  \textbf{9}(9), ~S285 (2007)

\bibitem{huang:07}
Huang, F.M., Zheludev, N., Chen, Y., Javier Garcia~de Abajo, F.: Focusing of light by a nanohole array. Applied Physics Letters  \textbf{90}(9) (2007)

\bibitem{jordan2020superresolution}
Jordan, A.N.: Superresolution using supergrowth and intensity contrast imaging. Quantum Studies: Mathematics and Foundations  \textbf{7}(3),  285--292 (2020)

\bibitem{jordan2024superphenomena}
Jordan, A.N., Aharonov, Y., Struppa, D.C., Colombo, F., Sabadini, I., Shushi, T., Tollaksen, J., Howell, J.C., Vamivakas, A.N.: Superphenomena for arbitrary quantum observables. Physical Review A  \textbf{110}(1),  012206 (2024)

\bibitem{PhysRevApplied.20.064046}
Jordan, A.N., Howell, J.C.: Fundamental limits on subwavelength range resolution. Phys. Rev. Appl.  \textbf{20},  064046 (Dec 2023). \doi{10.1103/PhysRevApplied.20.064046}, \url{https://link.aps.org/doi/10.1103/PhysRevApplied.20.064046}

\bibitem{PhysRevResearch.6.033341}
Jordan, A.N., Howell, J.C., Kempf, A., Zhang, S., White, D.: Optimal radar ranging pulse to resolve two reflectors. Phys. Rev. Res.  \textbf{6},  033341 (Sep 2024). \doi{10.1103/PhysRevResearch.6.033341}, \url{https://link.aps.org/doi/10.1103/PhysRevResearch.6.033341}

\bibitem{jordan2024quantum}
Jordan, A.N., Siddiqi, I.A.: Quantum Measurement: Theory and Practice. Cambridge University Press (2024)

\bibitem{karmakar2023beyond}
Karmakar, T., Jordan, A.N.: Beyond superoscillation: General theory of approximation with bandlimited functions. Journal of Physics A: Mathematical and Theoretical  \textbf{56}(49),  495204 (2023)

\bibitem{kempf2018four}
Kempf, A.: Four aspects of superoscillations. Quantum Studies: Mathematics and Foundations  \textbf{5}(3),  477--484 (2018)

\bibitem{kozawa:18}
Kozawa, Y., Matsunaga, D., Sato, S.: Superresolution imaging via superoscillation focusing of a radially polarized beam. Optica  \textbf{5}(2),  86--92 (2018)

\bibitem{kr2024experimental}
KR, S., Karmakar, T., Wadood, S., Jordan, A.N., Vamivakas, A.N.: Experimental realization of supergrowing fields. Physical Review Research  \textbf{6}(3),  L032043 (2024)

\bibitem{luo2020parametrization}
Luo, Q., Webb, K.J.: Parametrization of speckle intensity correlations over object position for coherent sensing and imaging in heavily scattering random media. Physical Review Research  \textbf{2}(3),  033148 (2020)

\bibitem{mccaul2023superoscillations}
McCaul, G., Peng, P., Martinez, M.O., Lindberg, D.R., Talbayev, D., Bondar, D.I.: Superoscillations deliver superspectroscopy. Physical Review Letters  \textbf{131}(15),  153803 (2023)

\bibitem{oliver1963sparkling}
Oliver, B.: Sparkling spots and random diffraction. Proceedings of the IEEE  \textbf{51}(1),  220--221 (1963)

\bibitem{redding2012speckle}
Redding, B., Choma, M.A., Cao, H.: Speckle-free laser imaging using random laser illumination. Nature photonics  \textbf{6}(6),  355--359 (2012)

\bibitem{redding2013all}
Redding, B., Popoff, S.M., Cao, H.: All-fiber spectrometer based on speckle pattern reconstruction. Optics express  \textbf{21}(5),  6584--6600 (2013)

\bibitem{Richard:59}
Richards, B., Wolf, E.: Electromagnetic diffraction in optical systems, ii. structure of the image field in an aplanatic system. Proceedings of the Royal Society of London. Series A. Mathematical and Physical Sciences  \textbf{253}(1274),  358--379 (1959)

\bibitem{Richards:59}
Richards, B., Wolf, E.: Electromagnetic diffraction in optical systems, ii. structure of the image field in an aplanatic system. Proceedings of the Royal Society of London. Series A. Mathematical and Physical Sciences  \textbf{253}(1274),  358--379 (1959)

\bibitem{rogers:12}
Rogers, E.T., Lindberg, J., Roy, T., Savo, S., Chad, J.E., Dennis, M.R., Zheludev, N.I.: A super-oscillatory lens optical microscope for subwavelength imaging. Nature materials  \textbf{11}(5),  432--435 (2012)

\bibitem{rogers:13needle}
Rogers, E.T., Savo, S., Lindberg, J., Roy, T., Dennis, M.R., Zheludev, N.I.: Super-oscillatory optical needle. Applied Physics Letters  \textbf{102}(3) (2013)

\bibitem{rosi2022theoretical}
Rosi, P., Venturi, F., Medici, G., Menozzi, C., Gazzadi, G.C., Rotunno, E., Frabboni, S., Balboni, R., Rezaee, M., Tavabi, A.H., et~al.: Theoretical and practical aspects of the design and production of synthetic holograms for transmission electron microscopy. Journal of Applied Physics  \textbf{131}(3) (2022)

\bibitem{schelkunoff1943mathematical}
Schelkunoff, S.A.: A mathematical theory of linear arrays. The Bell System Technical Journal  \textbf{22}(1),  80--107 (1943)

\bibitem{siegman:86}
Siegman, A.E.: Lasers. University science books (1986)

\bibitem{smith2020mathematical}
Smith, M.K., Gbur, G.: Mathematical method for designing superresolution lenses using superoscillations. Optics Letters  \textbf{45}(7),  1854--1857 (2020)

\bibitem{vsoda2020efficient}
{\v{S}}oda, B., Kempf, A.: Efficient method to create superoscillations with generic target behavior. Quantum Studies: Mathematics and Foundations  \textbf{7}(3),  347--353 (2020)

\bibitem{tsang:16}
Tsang, M., Nair, R., Lu, X.M.: Quantum theory of superresolution for two incoherent optical point sources. Physical Review X  \textbf{6}(3),  031033 (2016)

\bibitem{viteri2024supergrowth}
Viteri-Pflucker, V., Ryan, C.J., KR, S., Liang, K., Spiecker, D., Wadood, S., Jordan, A.N., Nick~Vamivakas, A.: Supergrowth in speckle patterns. Optics Letters  \textbf{50}(1),  137--140 (2024)

\bibitem{wang:08}
Wang, H., Shi, L., Lukyanchuk, B., Sheppard, C., Chong, C.T.: Creation of a needle of longitudinally polarized light in vacuum using binary optics. Nature photonics  \textbf{2}(8),  501--505 (2008)

\bibitem{wang:10}
Wang, T., Wang, X., Kuang, C., Hao, X., Liu, X.: Experimental verification of the far-field subwavelength focusing with multiple concentric nanorings. Applied Physics Letters  \textbf{97}(23) (2010)

\bibitem{white2024reconstructing}
White, D.D., Zhang, S., Soda, B., Kempf, A., Struppa, D.C., Jordan, A.N., Howell, J.C.: Reconstructing superoscillations buried deeply in noise. arXiv preprint arXiv:2410.05399  (2024)

\bibitem{Wu:18}
Wu, Z., Zhang, K., Zhang, S., Jin, Q., Wen, Z., Wang, L., Dai, L., Zhang, Z., Chen, H., Liang, G., Liu, Y., Chen, G.: Optimization-free approach for generating sub-diffraction quasi-non-diffracting beams. Opt. Express  \textbf{26}(13),  16585--16599 (Jun 2018)

\bibitem{zheludev:22}
Zheludev, N.I., Yuan, G.: Optical superoscillation technologies beyond the diffraction limit. Nature Reviews Physics  \textbf{4}(1),  16--32 (2022)

\bibitem{zhu2014arbitrary}
Zhu, L., Wang, J.: Arbitrary manipulation of spatial amplitude and phase using phase-only spatial light modulators. Scientific reports  \textbf{4}(1), ~7441 (2014)

\end{thebibliography}
\end{document}